\documentstyle[12pt,aaspp]{article}
\def\lsim{\raise2.90pt\hbox{$\scriptstyle
<$} \hspace{-6pt}\lower.5pt\hbox{$\scriptscriptstyle\sim$}\; }
\def\bi#1{\hbox{\boldmath{$#1$}}}
\def\ltsima{$\; \buildrel < \over \sim \;$}
\def\lsim{\lower.5ex\hbox{\ltsima}}
\def\gtsima{$\; \buildrel > \over \sim \;$}
\def\gsim{\lower.5ex\hbox{\gtsima}}

\begin{document}

\title{Weak Lensing Reconstruction and  
Power Spectrum Estimation: Minimum Variance Methods}
\author{Uro\v s Seljak}
\affil{ Harvard Smithsonian Center For Astrophysics, Cambridge, MA 02138 USA
}

\begin{abstract}
Large-scale structure distorts the images of background galaxies, which 
allows one to measure directly the projected 
distribution of dark matter in the universe and determine its power spectrum. 
Here we address 
the question of how to extract this information from the observations. 
We derive minimum variance estimators for projected density reconstruction
and its power spectrum and apply them to simulated data 
sets, showing that they give a good
agreement with the theoretical minimum variance expectations. 
The same estimator can also be applied to the cluster reconstruction,
where it remains a useful reconstruction technique, although it is
no longer optimal for every application. 
The method can be generalized to include
nonlinear cluster reconstruction and  
photometric information on redshifts of background
galaxies in the analysis.
We also address the question of how to obtain 
directly the 3-d power spectrum from the weak lensing data. 
We derive a minimum variance quadratic estimator, 
which maximizes the 
likelihood function for the 3-d power spectrum
and can be computed either from the measurements directly
or from the 2-d power spectrum. The estimator  
correctly propagates the errors and
provides a full correlation matrix of the estimates. It can 
be generalized to the case where 
redshift distribution depends on the galaxy photometric properties, which 
allows one to measure both the 3-d power spectrum and its time evolution.
\end{abstract}

\newpage

\section{Introduction}

The nature and distribution of dark matter is one of the great mysteries
of modern cosmology. Although most of the matter in the universe may
be dark, we can only directly observe baryons. By measuring statistical
properties of visible universe (galaxies and clusters) we cannot 
infer statistical properties of the underlying dark matter without further
assumptions. Typically one has to assume that light in some way 
traces mass, but this has often been questioned and models have 
been proposed where galaxy formation is a non-local (Heyl et al. 1995) 
or stochastic process (Dekel \& Lahav 1997).
Until recently the only two direct tracers of matter distribution on 
large scales with existing data 
were velocity flows and cosmic microwave background (CMB)
anisotropies. The former, while promising, still suffers from a small
number of galaxies with accurate distances and a 
number of inherent biases present in any of the analysis methods 
(\cite{sw}). In recent years CMB has emerged as
the most promising way to measure the dark matter distribution,
with the potential of measuring several cosmological parameters with 
a few percent accuracy over the next decade (\cite{jungman,bet,zss}). The main 
advantage of CMB over other methods is that it traces the universe
in linear regime, which makes the interpretation of the data relatively
straightforward. Another advantage is that it is relatively free of
systematic effects and foreground emission 
seems to be subdominant over a large range of angular scales.
However, CMB by itself cannot
determine all of the cosmological information (\cite{bet,zss}). 
In particular, it does not directly probe 
the matter power spectrum, which although related to the CMB power spectrum,
is sensitive to somewhat different physical processes and can provide 
valuable information by itself. For example,
massive neutrinos have only a minor effect on the CMB spectrum and it 
will be rather difficult to distinguish between different values
of neutrino mass from the CMB data alone. Their effect on the matter 
power spectrum is significantly more important and by measuring it one could
determine the value of the neutrino mass or set an upper limit.
For this reason it is important to investigate 
other direct tracers of matter, which may allow one a more direct 
determination of matter power spectrum.

It has long been recognized that gravitational
lensing offers the possibility of tracing the dark matter distribution
directly.
As light propagates through the universe it is deflected by the 
inhomogeneities along the line of sight and this causes a distortion in
the images of background galaxies. The distortions are small 
and for each individual galaxy one cannot separate them from an intrinsic
ellipticity of the galaxy. 
Only by averaging over several galaxies and
assuming that the ellipticities
of galaxies are uncorrelated can one
detect this so called weak lensing effect. The effect has already been 
detected in clusters, where the higher 
projected mass density and typical radial pattern simplify the detection
(see \cite{fm} for a review). The challenge for the future
is to detect the weak lensing away from clusters 
in the field, where the distortions are significantly smaller, 
but the rewards potentially higher (for some preliminary results see
\cite{vil}). 
Although the typical distortion in the field is smaller the survey
area can be larger (specially with the advent of
composite CCD cameras), which allows for a possibility of a statistical 
detection of structures on large scales through the two-point 
correlations, 
power spectrum or correlation 
function. Often this will be the most useful information anyways, since
on large scales, where the matter distribution is likely to be described as a 
gaussian random field, power spectrum contains all the information needed
for a statistical description of matter distribution. Power spectrum
is the starting point for the extraction of various cosmological parameters.
Cosmological model 
predictions for two-point statistics, both correlation functions
in real space and power spectrum
in Fourier space, have been investigated by a number of groups. The 
first calculations were those by \cite{Blandford91,Escude91} and 
\cite{Kaiser92} based on
the work by \cite{gunn67}. Recently they have been extended to a 
general Robertson-Walker universe by \cite{Bernardeau96}, Kaiser (1997),
Stebbins (1996), Seljak (1996) and
Jain \& Seljak (1997). The latter have also extended the calculations to a 
nonlinear regime using \cite{Hamilton91} mapping of the power spectrum. Most 
of these works have been theoretical and have only partially
addressed the issue of extraction of the signal from the data.  
Kaiser (1997) suggests a simple power spectrum estimator and shows
that it gives an unbiased estimate of a convolved power spectrum.
He also gives an estimate of the errors under the assumption that 
the data have gaussian distribution. 
Another way to obtain such a quantity is to transform the data to 
Fourier space first and then form a scalar quantity, 
as suggested by Kaiser (1992, 1997). This method
uses all the available information, but as we will show in this 
paper does not optimally weight
the data which leads to a loss of information.
\cite{sch97} propose a new
statistic called mass-aperture $M_{ap}$
(previously introduced by \cite{Kaiser94}
for estimating the mass of clusters) as a tool for estimating the 
power spectrum and higher order statistics. The statistic radially 
averages the tangential component of the shear.
It was originally
developed for clusters where for circular symmetry 
the shear pattern is indeed tangential. In the field this is no longer
the case and the motivation for using this statistic becomes less
obvious. One reason for introducing this statistic is to obtain 
a localized scalar quantity from which third and higher 
moments can be extracted.  Both of these papers do not address
the question of how efficiently the methods use the data
and how much information
is lost in the process of reducing the data to this form. 
We will show in this paper that there are methods which can formally
be shown to be optimal for the power spectrum reconstruction.
This is particularly important on large scales,
where sampling variance (finite
number of modes of a given wavelength in the survey box) is the
dominant source of error on the power spectrum.

Another topic that was not addressed in the context of weak lensing 
so far is the determination of 3-d power spectrum from 2-d data sets. 
One method proposed by \cite{be} and applied to APM galaxy survey 
is to compute 2-d power spectrum and
invert it to obtain 3-d power spectrum. The inversion procedure is
nonlinear and requires an iterative scheme. The main problem with this
analysis is the treatment of errors. On large scales the 
2-d estimators have a large sampling variance and unless this is 
properly taken into account they contribute too much weight to the 
3-d estimators and the inversion is unstable. 
We will show in this paper that there is
a well defined answer how to perform this procedure that formally 
minimizes the error and provides a correct error estimate for the 
3-d power spectrum. With weak lensing data one can extract
not only the power spectrum, but also its time evolution, if one 
uses photometric information to assign a distance to the background 
galaxies and we address how to extract this information as well.

Power spectrum is not the only information of interest when analyzing
such data sets.
On somewhat smaller scales higher order moments develop, 
which add additional
information on the statistical properties of the process and can 
independently constrain some of the cosmological parameters. For 
example, ratio of the third moment  
to the square of second moment (i.e. skewness) of convergence
depends in the second order perturbation theory only 
on the matter density $\Omega_m$ and the shape of the power spectrum, but 
not on its amplitude. This may allow one to measure directly the cosmic
density $\Omega_m$ (Bernardeau et al. 1997, \cite{js}). Because the measured 
shear is a tensorial quantity it has a vanishing skewness and cannot be 
used for this test. \cite{sch97} propose to use $M_{ap}$ as a statistic
on which to apply this test, but as mentioned above this statistic does
not make an optimal use of the data. We will show here that one can 
define a scalar quantity where all the information from the data is used
and provided all the statistical properties are properly included this
does not lead to a loss in information. One would also like to map large 
scale structures such as filaments, sheets and superclusters,  
which are of interest by themselves, both for cosmographical purposes
and for learning about the cosmological model from their properties. 
Again the question arises what is the optimal 
way to do this. We will show that for large scale structure this 
question has a well-defined answer, which moreover 
has a solution  
closely related to the optimal power spectrum reconstruction, so that 
both can be extracted within the same formalism.

In this paper we develop the so-called minimum variance linear 
methods for projected mass 
reconstruction, 2-d and 3-d
power spectrum estimation from the weak lensing data. 
We apply the methods to simulated data sets which demonstrate 
that the estimators are unbiased and reach the theoretical minimum 
variance limit. As such the estimators 
are superior to the estimators mentioned above
and should be used at least on large scales, 
where the nearly gaussian distribution guarantees their optimal 
performance. Although the methods developed here 
work best for reconstruction of LSS, they can be applied to the 
cluster reconstruction as well, where they still minimize the variance 
in the class of linear reconstruction methods. In this case they are
no longer optimal for every application and instead can be viewed only
as one of many possible linear filtering methods. Other linear or nonlinear 
methods may be better depending on the application one has in mind.

\section{Formalism}

Let us assume the observations consist of $N$ galaxies at angular 
positions $\bi{\theta}_i$ with 
measured ratios of short to long axis $b/a$ and position angle
of major axis $\phi$. We can introduce ellipticity $e=(1-b/a)/(1+b/a)$
from which we can form a two-component entity of observable quantities 
$\{e_1,e_2\}=\{e\cos 2\phi,e\sin 2\phi\}$.
We arrange all the data into a $2N$ component vector 
$\bi{e}=\{e_1(\bi{\theta}_i),e_2(\bi{\theta}_i)\}(i=1,...,N)$. 
Because of gravitational lensing the
``true'' surface brightness (i.e. the surface brightness one would see 
in the absence of any lensing) is mapped into the observed one,
$I_{{\bf obs}}(\bi{ \theta})=I_{{\bf true}}(\bi{ \theta} +\delta \bi{\theta})$,
where $\delta \bi{\theta}$ is
the angular deflection of a photon caused by intervening mass 
distribution.
This induces ellipticity correlations between the galaxies, information 
on which is contained
in the symmetric deformation tensor,
\begin{eqnarray}
\Phi_{ij} &\equiv & {\partial \delta \theta_i \over \partial \theta_j}
=- 2 \int_0^{\chi_0}
g(\chi)\nabla_i \nabla_j \phi(\chi) d\chi \nonumber \\
g(\chi)&=&r(\chi) \int_\chi^{\chi_0}
{r(\chi' -\chi) \over r(\chi')}W(\chi')d\chi'\ .
\label{shear}
\end{eqnarray}
Here $\phi$ is the gravitational potential and $\chi$ is the radial
comoving distance with $\chi_0$ being the horizon distance.
The radial distribution of background galaxies is described with
the normalized distribution $W(\chi')$. 
The comoving angular distance $r(\chi)$ introduced above
can be expressed in terms of $\chi$ as 
\begin{eqnarray}
r(\chi)=
\left\{ \begin{array}{ll} K^{-1/2}\sin K^{1/2}\chi,\ K>0\\
\chi, \ K=0\\
(-K)^{-1/2}\sinh (-K)^{1/2}\chi,\ K<0\\
\end{array}
\right.
\label{rchi}
\end{eqnarray}
Curvature $K$ can be expressed as 
$K=(\Omega_{\Lambda }+\Omega_{m}-1)H_0^2$, where $H_0$ is the Hubble 
constant and $\Omega_{\Lambda }$, $\Omega_{m}$ 
the vacuum and matter densities, respectively.
Deformation tensor can be decomposed into its trace component 
convergence $\kappa$
and two traceless components of the shear $\gamma_1,\gamma_2$,
$\Phi_{11}=1-\kappa-\gamma_1$, $\Phi_{22}=1-\kappa+\gamma_1$ and
$\Phi_{12}=\Phi_{21}=-\gamma_2$. The
observable ellipticities are related to these via the expression
$e_i=\gamma_i /(1-\kappa)$ (\cite{kochanek,Kaiser95,ss}; we are assuming
that there are no critical lines where $|\det \Phi |=0$ and between
which this relation changes, see section 4) and so only a 
combination of $\kappa$ and $\gamma_i$ can be determined. However, since
the distortions induced by LSS are small, i.e. $\kappa \ll 1$, one can 
in the first approximation
ignore $1-\kappa$ in the above expression and then shear becomes 
directly expressed
in terms of the observed ellipticity. Nonlinear
corrections are addressed in more detail in section 4.

Although there are 3 components of the deformation tensor they are 
in fact not independent, because they can all be 
expressed as a derivative
of a single deflection potential  
(equation \ref{shear}). The quantity that we are mostly
interested in is the convergence or surface mass density 
$\kappa$, which is related to the
projected mass distribution over the window given in equation (\ref{shear}). 
We would like to reconstruct it and compute its power spectrum 
from the observables $\bi{e}$.  
The relation between the shear and 
convergence is most easily expressed in Fourier space (throughout
the paper we will assume that the scales are sufficiently small 
for curvature of the sky to be unimportant and so Fourier analysis 
to be adequate; see \cite{stebbins96} for a generalization 
of this to all-sky measurements), so we first decompose shear and
convergence into a Fourier series,
\begin{equation}
\kappa(\bi{\theta})=\sum_{\bi{l}} e^{i\bi{l}\cdot\bi{\theta}} 
\tilde{\kappa}(\bi{l}) 
\end{equation}
where $\tilde{\kappa}(\bi{l})$ is the power per mode and the
sum is taken in the limit where the spacing between the modes
$\Delta l \rightarrow 0$.
The two shear components can be similarly expanded and the relation 
between them and convergence in Fourier space is (Kaiser 1992)
\begin{equation}
\tilde{\gamma}_1(\bi{l})=\tilde{\kappa}(\bi{l})\cos 2\phi_l\,\,\,,\,\,\,
\tilde{\gamma}_2(\bi{l})=\tilde{\kappa}(\bi{l})\sin 2\phi_l ,
\label{shearkappa}
\end{equation}
where $\phi_l$ is the direction angle of $\bi{l}$ mode. The data can be 
modelled in the form 
\begin{equation}
\bi{e}=\bi{R}\tilde{\bi{\kappa}} + \bi{n},
\label{e}
\end{equation}
where $\tilde{\bi{\kappa}}(\bi{l})$ 
is the underlying field in Fourier space, $\bi{n}$ is the noise vector,
$\bi{R}$ is a $2N \times M$ response matrix of the form 
$R(\bi{\theta})=e^{i\bi{l}\cdot\bi{\theta}}
\{\cos 2\phi_l ,\sin 2\phi_l \}$
for the two components of $\bi{e}$ at position $\bi{\theta}$ and 
$M$ is the number of modes we use in the model. 
Instead of 
working with the complex response matrix $\bi{R}$ and the complex field
$\tilde{\bi{\kappa}}$ (which satisfies the condition $\tilde{\kappa}(\bi{l})=
\tilde{\kappa}^*(-\bi{l})$) we can also work with
real valued response matrix
by replacing $e^{\pm i\bi{l}\cdot\bi{\theta}}$ with $\sqrt{2}\cos(\bi{l}\cdot\bi{\theta})$
and $\sqrt{2}\sin(\bi{l}\cdot\bi{\theta})$ and replacing the complex
$\tilde{\kappa}(\bi{l})$ with two real components.
Underlying field always has an infinite number of coefficients, but for
computational reasons only a finite number of these can be included.
The required number depends on the size of the survey and signal to noise
ratio.
The spacing between the modes $\Delta l$ is determined by the size of 
the survey $L$,
$\Delta l = 2\pi/L$, since one cannot resolve the modes more finely
spaced than this. One way to see this is to 
embed the survey area into a box of size $L^2$ and impose periodic 
boundary conditions on the modes to make them orthogonal. 
These modes are spaced 
by $\Delta l$ and form a complete and orthonormal system
in the box, so
that any distribution can be expanded into this series. Such an 
expansion will impose periodic boundary conditions on the data. 
To avoid spurios effects from periodic boundary condition
it is necessary 
to have the box somewhat larger than the survey and typically 
20\% zero padding on each side works well (see section 4). 
Even better is 
to expand in a box of twice the linear size, which will completely 
eliminate the boundary condition problems but with an increase in the 
number of modes.
The upper cutoff in the expansion is determined by the signal to noise ratio:
because of the noise 
the data do not contain information about the structure on 
very small scales and do not need to be modelled. One can always
include more modes than necessary or embed the survey into a larger box, 
so in this sense their actual 
number is not very important, except for computational reasons. 

In addition to the contribution from 
LSS each measurement also has a noise contribution $n_i$, which is 
a combination of intrinsic ellipticity of the galaxy 
(dominant for bright and/or large
sources)
and the measurement error (significant for faint and/or small sources). We will
assume that the noise covariance matrix is diagonal and given by 
$\bi{N}=\langle \bi{nn}^{\dag}\rangle=\sigma^2_i \delta_{ij}$, where 
$^{\dag}$ is the transpose (and conjugate for complex fields) of the vector.
Measurement error 
can differ from 
one galaxy to another and intrinsic rms ellipticities can vary with 
galaxy types and/or redshift, 
so in principle the variances $\sigma^2_i$ for different galaxies 
do not have  
to be equal and one can include this information if necessary. 
For a given galaxy the errors on $e_1$, $e_2$
are assumed equal and uncorrelated. 

The fluctuations in the underlying field $\tilde{\bi{\kappa}}$ 
are parametrized with a diagonal covariance matrix $\tilde{\bi{S}}$,
$\tilde{S}_{\bi{ll'}}=\langle \tilde{\kappa}_{\bi{l}}\tilde{\kappa}_{\bi{l'}}\rangle=
P(l)\delta_{\bi{ll'}}(\Delta l)^2$, where $P(l)$ is the convergence 
power spectrum at $l$, amplitude of $\bi{l}$ mode. 
It can be expressed as an integral over the matter 
density power spectrum (Kaiser 1992, \cite{Kaiser97,js})
\begin{equation}
P(l)={9 \pi\over 2}\ \Omega_m^2\ \, H_0^2\,
\int_0^{\chi_0}\ {g^2(\chi) \over r^2(\chi)}\
a^{-2}(\chi)\ P_\delta[k=l / r(\chi),\chi]d\chi,
\label{ps}
\end{equation}
where $P_\delta(k,\chi)$ is the density power spectrum at conformal 
time $\tau=\chi_0-
\chi$ and $a=(1+z)^{-1}$ the expansion 
factor, related to the radial distance $\chi$ via the relation 
$da/d\chi=H_0[\Omega_{m}a+
\Omega_{\Lambda } a^4 + (1-\Omega_{\Lambda }-\Omega_{m})a^2]^{1/2}$.
We will also need the correlation
matrix for the data $\bi{C}=\langle \bi{e}\bi{e}^{\dag} \rangle$. It can
be computed from the power spectrum as
\begin{equation}
\langle e_{\alpha}(\bi{\theta}_i)e_{\beta}(\bi{\theta}_j) \rangle=
\pi\int ldlP(l)A_{\alpha\beta}(l\vert \bi{\theta}_i-\bi{\theta}_j \vert)
+\sigma^2_i\delta_{ij}, 
\label{corr}
\end{equation}
where 
\begin{eqnarray}
A_{11}(x)&=&J_0(x)+\cos(4\phi)J_4(x)\nonumber \\
A_{22}(x)&=&J_0(x)-\cos(4\phi)J_4(x)\nonumber \\
A_{12}(x)&=&A_{21}(x)=\sin(4\phi)J_4(x).
\label{corrang}
\end{eqnarray}
Here $\phi$ is the direction angle of $\bi{\theta}_i-\bi{\theta}_j$ and
$J_n(x)$ are the Bessel functions of order $n$.

\subsection{Optimal Filtering}
Reconstruction techniques for 
LSS and CMB have been extensively studied (\cite{bunn,zar}). 
A particularly 
simple nonparametric technique is optimal or Wiener filtering (WF), which  
minimizes the variance in the class of linear estimators 
(\cite{rp,zar}). When the data are gaussian distributed it 
coincides with
the maximum posterior probability estimator and so is optimal in this 
limit. As such it should be 
the ideal method when applied to the 
LSS reconstruction from weak lensing data,
where the deviations from gaussianity are either 
small or negligible. Another advantage of WF is that it provides an 
analytic expression for the error covariance matrix, which can 
quantify which structures are statistically significant.
One problem of WF is that it requires
as input the power spectrum of the underlying field and this is often 
not known in advance. In the absence of any external information it 
has to be obtained from the data itself. We will show that
one can obtain a power spectrum estimator 
directly from the WF reconstruction, thus providing 
a self-consistent treatment of WF without the need of having two 
separate analysis for the two problems (\cite{seljak97b}). This 
power spectrum is in fact the optimal one in the sense of being 
minimum variance among all the estimators. 
Here we will 
review this formalism, modifying it where necessary to the case of
weak lensing.

Let us begin with the reconstruction of surface density 
in Fourier space $\tilde{\bi{\kappa}}$. We can subsequently make 
a real space map by transformation $\bi{\kappa}=\bi{R}_{\kappa}
\tilde{\bi{\kappa}}$, where $R_{\kappa}(\bi{\theta})=
e^{i\bi{l}\cdot\bi{\theta}}$,
with $\bi{\theta}$ the angular position where we want the value
of the reconstructed field.
We will require that the
estimated field $\hat{\tilde{\bi{\kappa}}}$ 
is a linear function of the data, $\hat{\tilde{\bi{\kappa}}}=
\bi{\Phi d}$, where $\bi{\Phi}$ is a $2N \times M$ matrix. Then one can 
minimize the variance of the residual 
\begin{equation}
\langle (\tilde{\bi{\kappa}}-\hat{\tilde{\bi{\kappa}}})
(\tilde{\bi{\kappa}}-\hat{\tilde{\bi{\kappa}}})^{\dag} \rangle
\label{res}
\end{equation}
with respect to $\bi{\Phi}$. This gives
the WF estimator of the underlying field,
\begin{equation}
\hat{\tilde{\bi{\kappa}}}=\langle \tilde{\bi{\kappa}}\bi{e}^{\dag}\rangle \langle \bi{ee}^{\dag}
\rangle^{-1}=\tilde{\bi{S}}\bi{R}^{\dag} \bi{C}^{-1}\bi{e}.
\label{wf}
\end{equation}
The variance of residuals is given by
\begin{equation}
\langle \tilde{\bi{r}}\tilde{\bi{r}}^{\dag} \rangle =\langle 
(\tilde{\bi{\kappa}}-\hat{\tilde{\bi{\kappa}}})
(\tilde{\bi{\kappa}}-\hat{\tilde{\bi{\kappa}}})^{\dag}
\rangle=
\tilde{\bi{S}}-\tilde{\bi{S}}\bi{R}^{\dag}\bi{ C}^{-1} \bi{R}\tilde{\bi{S}}
.
\label{var}
\end{equation}
For gaussian random fields Wiener filtering coincides with 
maximum probability  
estimator (\cite{zar}), which maximizes the likelihood function (or 
posterior probability in Bayesian language)
and so it is optimal. 
This is because WF only uses information
on the mean and variance of statistical distribution, which 
completely characterize gaussian random fields.
Since WF basically multiplies the data with signal/(signal+noise)
one can see that the modes with low signal to noise 
are being filtered out and replaced with zero. Only the modes with 
significant signal to noise
will be kept in the reconstruction, thus providing the most conservative
estimate of the field. For gaussian random 
fields this is in fact the optimal reconstruction. 
By using only information on the mean and variance 
it may be less than optimal when applied 
to a strongly nongaussian field,  
although even in such situations WF still minimizes the variance
as defined in equation (\ref{res})
in the class of linear estimators.
For cluster 
reconstruction from weak lensing there have been proposed
several methods that
are linear in the data (see e.g. \cite{sk} and references therein), 
but since WF explicitly 
minimizes the variance in equation (\ref{res}) it is in 
this sense optimal in this class of reconstruction 
methods. We will discuss this in more detail in section 4, where 
we show that in such applications minimizing 
the variance may not be the only way to define the best image
reconstruction and other reconstruction methods may be better 
depending on the application.
Note that WF requires as input the knowledge of power spectrum $P(l)$
to compute $\tilde{\bi{S}}$ and $\bi{C}$. To make the method self-consistent 
we therefore need a power spectrum estimator as well. We turn to this
next.

\subsection{2-d minimum variance power spectrum estimator}

Optimal power spectrum reconstruction from a 
set of noisy and incomplete observations has 
been investigate extensively in the field of 
large scale structure (LSS) and cosmic microwave background (CMB)
anisotropies (\cite{kbj,hama,teg,gorski}), inspired by the 
existing and forthcoming large data sets (COBE, MAP, Planck, SDSS, 2DF etc.).
The amount of information loss when analyzing the data can be 
quantified with the use of information theory, which allows one to define 
the requirements for an optimal estimator (\cite{tth}).
One can define the Fisher information 
matrix as the ensemble average of the 
matrix of second derivatives of the (minus) 
log-likelihood function with respect to the parameters we wish to estimate.
Its inverse provides a minimum bound for 
covariance matrix of the parameters, known as the 
Cram\' er-Rao inequality (e.g. Kendall and Stuart 1969). Power 
spectrum estimator that is quadratic in the data
has been proposed that reaches this theoretical limit directly 
(\cite{hama,teg,kbj}) and it has been shown that it gives  
equivalent results to the maximum likelihood analysis (\cite{kbj}).
Under the gaussian assumption for the data the method also provides
the error matrix for the estimators, which includes contributions 
from the sampling variance, noise and aliasing.
The main shortcoming of this estimator is that it is still 
computationally expensive for large data sets, but this can be 
significantly reduced by transforming the data to a signal eigenmode
basis (\cite{seljak97b,spergel}) 
or by using various approximations. 

The power spectrum $P(l)$ is a continuous function of $l$, but we
only sample it in a finite number of modes, so we can only estimate 
it at discrete values of $l$. Moreover, we can group together the 
estimates from modes that are nearby in $l$ to reduce the scatter in 
the estimator. The spacing in $l$ can be $\Delta l$ or 
larger, depending on the amount of information in the data and on the 
smoothness of the power spectrum.
In the following we will number these with $l$ and 
denote the corresponding power per mode 
estimator with $\hat{\Theta}_l=\hat{P}(l)\Delta^2 l$, 
which can be put into 
a vector $\hat{\bi{\Theta}}$. 
We can introduce a
projection matrix $\bi{\Pi}_l$, which consists of ones along the diagonal
corresponding to the modes contributing to $l$-th parameter and zeros
everywhere else.
If we approximate the integrals in the correlation function (equation 
\ref{corr}) with the sum over the discrete modes used in the expansion
we can write 
\begin{equation}
\bi{C}=\sum_l \bi{\Pi}_l\bi{R} \tilde{S}_l \bi{R}^{\dag} \bi{\Pi}_l
\equiv \sum_l \tilde{S}_l \bi{Q}_l,
\label{q}
\end{equation}
where we introduced $\bi{Q}_l \equiv 
{\partial \bi{C} \over \partial \Theta_l}=\bi{\Pi}_l\bi{R}\bi{R}^{\dag}\bi{\Pi}_l$. 

To derive the minimum variance power spectrum 
estimator we write the likelihood function for the data
\begin{equation}
L(\bi{e} \vert \bi{\Theta})=(2\pi)^{-N/2} \det(\bi{C})^{-1/2} \exp(-{1 \over
 2}
\bi{e}^{\dag} \bi{C}^{-1}\bi{e})
\label{lik}
\end{equation}
and following \cite{kbj} we expand it to second order in parameters 
$\bi{\Theta}$,
\begin{equation}
\ln L(\bi{\Theta}+\delta \bi{\Theta})= \ln L(\bi{\Theta})+\sum_l
{\partial \ln L(\bi{\Theta}) \over \partial \Theta_l}\delta \Theta_l
+ {1 \over 2} \sum_{ll'} {\partial^2 \ln L(\bi{\Theta}) \over 
\partial \Theta_l\partial \Theta_l'}\delta \Theta_l\delta \Theta_l'.
\end{equation}
The derivatives are given by 
\begin{eqnarray}
-2{\partial \ln L(\bi{\Theta}) \over \partial \Theta_l} &=&
tr(\bi{e}^{\dag} \bi{C}^{-1} \bi{Q}{_l}\bi{C}^{-1}\bi{e} -\bi{C}^{-1} 
\bi{Q}_l) \nonumber \\
-{\partial^2 \ln L(\bi{\Theta}) \over 
\partial \Theta_l\partial \Theta_l'}&=&\bi{e}^{\dag} \bi{C}^{-1} 
\bi{Q}_l\bi{C}^{-1}\bi{Q}_{l'}\bi{C}^{-1}\bi{e}-
{1 \over 2} tr(\bi{C}^{-1} \bi{Q}_l\bi{C}^{-1}\bi{Q}_{l'}).
\end{eqnarray}
Ensemble average of the second expression is the Fisher matrix
\begin{equation}
F_{ll'}\equiv \Big\langle -{\partial^2 \ln L(\bi{\Theta}) \over
\partial \Theta_l\partial \Theta_l'} \Big\rangle = 
{1 \over 2} tr(\bi{C}^{-1} \bi{Q}_l\bi{C}^{-1}\bi{Q}_{l'}).
\label{fisher}
\end{equation}
At the maximum likelihood value the first derivative of the likelihood
function vanishes, so we use Newton-Raphson method to find the zero
of the derivative. This leads to a 
quadratic estimator, which is to be solved iteratively, 
\begin{equation}
\delta \Theta_l= {1 \over 2} \bi{F}_{ll'}^{-1}tr[(
\bi{e}^{\dag}\bi{e}-\bi{C})\bi{C}^{-1}\bi{Q}_{l'}\bi{C}^{-1} ].
\end{equation}
Once the power spectrum has been determined one can 
express the estimator directly
in terms of WF coefficients using equation (\ref{q}), taking the 
initial estimate to be zero (and the corresponding correlation 
matrix pure noise). This leads to the minimum variance quadratic 
estimator 
(\cite{hama,teg,seljak97b})
\begin{eqnarray}
\hat{\Theta}_l&=&{1 \over 2}\sum_{l'}F^{-1}_{ll'}[\bi{e}^{\dag}
\bi{C}^{-1}\bi{Q}_{l'}\bi{C}^{-1}\bi{e}-b_{l'}]=
{1 \over 2}\sum_{l'}F^{-1}_{ll'}[
\bi{y}^{\dag} \bi{\Pi}_{l'}
\bi{y}-b_{l'}] \nonumber \\
\bi{y}& \equiv &\tilde{\bi{S}}^{-1}\hat{\tilde{\bi{\kappa}}}=
\bi{R}^{\dag} \bi{C}^{-1} \bi{\kappa} \nonumber \\
b_l&=&tr[\bi{C}^{-1}\bi{Q}_l\bi{C}^{-1} (\bi{N+S_b})]=
tr[\bi{C}^{-1}\bi{R}\bi{\Pi}_l\bi{R}^{\dag} \bi{C}^{-1}\bi{(N+S_b})]
\nonumber \\
F_{ll'}&=& {1 \over 2} 
\sum_{\bi{l}} \sum_{\bi{l}'} 
|\bi{\Pi}_l\bi{R}^{\dag} \bi{C}^{-1} \bi{R}\bi{\Pi}_{l'}|_{\bi{ll}'}
^2.
\label{wffll}
\end{eqnarray}
We included the additional noise term
$\bi{S_b}$ in equation (\ref{wffll}), which is the correlation 
matrix for the modes that are not being estimated and which 
alias power into the modes we do estimate (\cite{seljak97b}).
The estimator above not only improves upon
the naive power spectrum estimation obtained by a simple average of 
Wiener filtered modes (which is always biased towards low values),
but is also the best possible estimator, since it gives equivalent
results to the maximum likelihood analysis and the 
covariance matrix of the estimates is given by the Fisher matrix,
which by Cram\' er-Rao inequality is the best one can do. To 
compute the estimator one has first to multiply the data with the
inverse of correlation matrix. This
first step is identical to WF and is just a generalization of the 
inverse noise weighting of the data. If a given measurement has a	
large noise or is strongly correlated with other measurements 
then it does not add new 
information and should be downweighted. 
Once this has been done subsequent operations do not require 
any additional weighting:  
to compute the power spectrum estimator one Fourier transforms the data
and adds the squares of all the modes  
contributing to the $l$-th parameter in 
the spectrum. There is no need to put additional weighting on the 
modes since correct weighting has already been obtained by the first operation.
For the estimator to be unbiased we need to subtract the noise and aliasing bias $b_l$ and compute the window function 
$\bi{F}$. 
The last step is to deconvolve the estimators with $\bi{F}^{-1}$, 
which may
not be recommended if the matrix is nearly singular, since 
the inversion may not be stable. Instead
one can quote convolved
estimators $N_l\bi{F}\hat{\bi{\Theta}}$,
where $N_l=(\sum_{l'}F_{ll'})^{-1}$.
The filtering function
$N_l\bi{F}$ is a bell shaped function in $l'$ for a given $l$
and its width gives the spectral resolution at that amplitude.
Under the gaussian approximation for the data the 
covariance matrix for this estimator is given by $N_lN_{l'}F_{ll'}$.
Note that only the first step, multiplying the data with $\bi{C}^{-1}$
differs from the uniform weighting 
power spectrum estimator proposed by Kaiser (1997).
This suggests that one can use an approximate form for $\bi{C}^{-1}$,
which is easier to invert, and still obtain an unbiased estimator. The
closer this matrix is to $\bi{C}^{-1}$ the closer will the estimate be to 
the minimum variance one.

The most expensive operation in computing WF is inverting 
$\bi{C}$ and once this inversion is obtained one can 
compute $\bi{b}$ and $\bi{F}$ by taking matrix products, 
so computing the power spectrum from 
WF estimators is not significantly more expensive than computing WF itself.
This inversion is O($N^3$) operations, if one uses Cholesky decomposition 
and becomes computationally too expensive when the data set becomes
too large ($N\gsim 10^4$). Since the shear measurements are not 
independent (in the sense that both components can be derived from a 
single scalar field) one can reduce the size of the matrix to be 
inverted by first transforming the data multiplied with the inverse of
the noise matrix into the Fourier space,
$\tilde{\bi{e}}=
\bi{R}^{\dag} \bi{N}^{-1} \bi{e}$. The modes $\tilde{\bi{\kappa}}$
we wish to estimate are related to these through the relation
\begin{equation}
\tilde{\bi{e}}=\bi{R}^{\dag} \bi{N}^{-1}\bi{R}\tilde{\bi{\kappa}}
\equiv \tilde{\bi{N}}^{-1} \tilde{\bi{\kappa}}
\label{tilkap}
\end{equation}
and their covariance matrix is given by 
\begin{equation}
\tilde{\bi{C}}=\langle \tilde{\bi{\kappa}}
\tilde{\bi{\kappa}}^{\dag}\rangle =\tilde{\bi{S}}+\tilde{\bi{N}}.
\label{covtilkap}
\end{equation}
If we use these modes in the likelihood function (equation \ref{lik})
and in the corresponding quadratic estimator (equation \ref{wffll}) we
reduce the matrix by a factor of two (and the computational cost by a factor
of eight), even if we use the same number of modes
as the number of galaxies. Moreover, since weak lensing
measurements have a low signal to noise ratio  
we need to model far fewer modes than we have galaxies, both 
of which suggest to use the Fourier space to perform the 
expensive matrix inversion and multiplication.
To avoid having to invert both $\tilde{\bi{N}}^{-1}$
and $\tilde{\bi{C}}$ we can use the relation 
\begin{equation}
\tilde{\bi{C}}^{-1}=\tilde{\bi{S}}^{-1}\bi{D}^{-1}\tilde{\bi{N}}^{-1}\, ,\,
\,\,
\bi{D} \equiv \tilde{\bi{S}}^{-1}+\tilde{\bi{S}}^{-1}\tilde{\bi{N}}(\bi{R}^{\dag} 
\bi{N}^{-1}\bi{S_b}\bi{ N}^{-1} \bi{R})+
\tilde{\bi{N}}^{-1},
\end{equation}
in which case if aliasing can be ignored only the matrix $\bi{D}$ 
needs to be inverted (otherwise we still have to invert $\tilde{\bi{N}}^{-1}$ as well).
Using this transformation the analogous 
expressions to equations (\ref{wffll})
are (\cite{seljak97b}),
\begin{eqnarray}
\tilde{\bi{S}}\bi{y}&=&
\hat{\tilde{\bi{\kappa}}}= \bi{D}^{-1} \tilde{\bi{e}} \nonumber \\
b_l&=&tr[\tilde{\bi{C}}^{-1}\tilde{\bi{Q}}_l\tilde{\bi{C}^{-1}}(\tilde{\bi{N}}
+\tilde{\bi{S}}_b)] \nonumber \\
F_{ll'}&=&{1 \over 2} tr[\tilde{\bi{C}}^{-1}\tilde{\bi{Q}}_l
\tilde{\bi{C}}^{-1}\tilde{\bi{Q}}_{l'} ] 
\label{signal}
\end{eqnarray}
The role of the correlation matrix $\bi{C}$ in real space has 
now been replaced by $\bi{D}$ (or $\tilde{\bi{C}}$), 
which has dimensions $M \times M$. 
If $M$ is significantly smaller than $N$ then a substantial saving in 
computational time can be achieved by having to invert and multiply
smaller matrices. 

For very large $M$ even performing these matrix manipulations
becomes computationally too
expensive, so it is worth exploring approximations to the above 
expressions, which would allow one to compute them more rapidly. 
If the survey is compact and the sampling relatively uniform 
then on small scales there will be little mixing between 
the modes and the geometry of the survey will not be important,
so to estimate the power at a given amplitude one 
can approximate the spectrum around it as flat (white noise). 
For such
a power spectrum the correlation matrix $\bi{C}$ in equation (\ref{corr}) is
diagonal,
$C_{ij}=(\sigma^2+\sigma_s^2
)\delta_{ij}$, where $\sigma^2$ is the noise variance for each galaxy 
and $\sigma_s^2=(2\pi)^2\Theta_l\bar{n}(\bi{\theta})$ 
is the theoretical variance, $\bar{n}(\bi{\theta})$ being the mean
density of galaxies at location $\bi{\theta}$.
This matrix 
can be inverted trivially in real space and $(\bi{C}^{-1}\bi{e})_i=
e_i/(\sigma^2+\sigma_s^2)$ 
is a simple inverse weighting of the data, where the weight 
consists now of both noise and signal variance.  
WF estimator is give by equation (\ref{wf})
with the variance of residuals
\begin{equation}
\langle r_{\bi{l}}r^*_{\bi{l'}} \rangle =
\Theta_l[\delta_{\bi{l}\bi{l'}}-\Theta_l\tilde{\bi{C}}^{-1}(\bi{l}-\bi{l'})].
\end{equation}
The bias and Fisher matrix are given by
\begin{eqnarray}
b_l&=&M_l \sum_i {\sigma^2 \over (\sigma^2+\sigma_s^2)^2}
\nonumber \\
F_{ll'}&=&\sum_{\bi{l}}\sum_{\bi{l'}} 
|\tilde{\bi{C}}^{-1}(\bi{l-l'})|^2.
\end{eqnarray}
Here $\bi{l}$, $\bi{l}'$ are the wavevectors corresponding 
to parameters $\Theta_l$ and $\Theta_{l'}$, respectively, and 
$\tilde{\bi{C}}^{-1}$ is a Fourier transform of $(\sigma^2+
\sigma_s^2)^{-1}$.
We see that
a complete solution to the problem can be obtained 
without performing any matrix 
inversion. Moreover, one can take advantage of Fast Fourier Transform 
(FFT) to make all the operations O($N \ln N$).
If the galaxy density does not vary across the field 
then this estimator agrees with the
uniform weighting of the data, as proposed by Kaiser (1997).
For a compact survey and for wavelengths smaller than the size of the survey 
this estimator will indeed be close to optimal (\cite{hama}). However,
for wavelengths comparable to the size of the survey or for sparse surveys
it is better to use the exact form of the estimator  in equations 
(\ref{wffll}).
The reason for this is that the multiplication with the inverse of the
correlation matrix apodizes the kernel of the window function and reduces
the mixing between the modes. The 
amount of apodizing depends on the actual power spectrum. 
If the power spectrum
is white noise then all the data points are uncorrelated and there is 
no apodization
even for wavelengths comparable to the size of the box.
If however the power spectrum has a lot of long wavelength power 
then uniform weighting introduces edge effects and one 
has to suppress these by tapering the edges. Multiplying the data with the
inverse of correlation function does this in an optimal way. 

\subsection{3-d  minimum variance power spectrum estimator}
So far we discussed the reconstruction of the 2-d power spectrum. This is 
the first method one should attempt, 
since the data are given in the form of a
2-d distribution and to look for a detection one should first look at
the 2-d power spectrum. Moreover, 2-d power spectrum 
contains all the information about the data, so by reducing the data
to this form no information has been lost (but see 
next section for the more general case when the 
redshift information of the background galaxies is included). 
However, ultimately the underlying quantity we wish to determine
is the 3-d power spectrum. For any given 3-d power spectrum
one can always predict the 2-d power spectrum by performing the 
integral in equation (\ref{ps}), but it is also useful to reconstruct 
the 3-d power spectrum directly to facilitate the comparison 
with cosmological models. This will depend on the assumed 
curvature and cosmological constant, because one has 
to translate observed angle to a comoving scale and 
include the effects of power spectrum changing with time, both in 
linear and nonlinear regime. It will also depend on the assumed 
distribution of background galaxies. 3-d power spectrum estimation is thus
more model dependent than the 2-d power spectrum itself, but is also 
closer to what we actually want to determine. 
Previous attempts in this direction
have used nonlinear inversion of the 2-d power spectrum in the APM survey
(\cite{be,gb}), but the treatment of the errors was not discussed and
because the 2-d power spectrum estimates have varying errors this can 
have a large effect on 
the reconstructed 3-d power spectrum. This is particularly important 
on large scales, where the modes have large sampling variance and have 
to be correspondingly downweighted. Here we derive 
a quadratic estimator for the 3-d power spectrum expanding the 
likelihood around its maximum following 
the procedure leading to the minimum variance power spectrum in 2-d, 
which guarantees the estimator will 
achieve minimum variance bound and provide a 
full covariance matrix for the estimates at the same time. 
This procedure could be useful whenever
one wishes to translate 1-d or 2-d observations to a 3-d power spectrum. 

The easiest way to derive the estimator is to
work with the data in the form of $\tilde{\bi{\kappa}}$ defined in 
equation (\ref{tilkap}). The corresponding covariance matrix is given in 
equation (\ref{covtilkap}) with the signal part $\tilde{\bi{S}}$ being diagonal.
Instead of determining the power spectrum $P_{\delta}(k)$ directly
we will determine its deviations from a fiducial power spectrum 
$P^0_{\delta}(k)$,
\begin{equation}
P_{\delta}(k)\Delta \ln k=P^0_{\delta}(k)T_k.
\end{equation}
We divided $P_{\delta}(k)/P^0_{\delta}(k)$ 
into bins of width $\Delta \ln k$ (we
choose to model the power spectrum with  
logarithmically spaced estimates) and denote each contribution $T_k$.
This will facilitate the interpretation of the window function and
will give a better estimate if we choose not to deconvolve the 
power spectrum with the window in the end (the same should also 
be used in estimating the 2-d power spectrum, although when there
is little mixing between the modes it may not be very important, as 
long as the power spectrum is not strongly varying across the 
bin width $\Delta l$).
Density power 
spectrum grows with time and we will parametrize it with 
$P^0_{\delta}(k,\chi)=P^0_{\delta }(k)a^2(\chi)F^2(\chi)$, where 
$P^0_{\delta }(k)$ is the power spectrum today and $F^2(\chi)$ 
is the growth factor of gravitational potential normalized to unity today, 
which can be expressed
as (Lahav et al. 1991)
\begin{eqnarray}
F(\chi) = 2.5\, \Omega_m\, a^{-1}\, (xf+1.5\Omega_m a^{-1}+
\Omega_K)^{-1} \nonumber\\
x=1+\Omega_m(a^{-1}-1)+\Omega_\Lambda(a^2-1)\, \,
;\, f=\left({\Omega_m \over ax} \right)^{0.6}.
\label{f}
\end{eqnarray}
For a flat $\Omega_m=1$ model gravitational potential does not change 
with time and $F$ is unity at all times. The expression above
is further modified in the nonlinear regime, where 
growth factor becomes a function of both $\chi$ and $k$ (Hamilton et al. 1991, \cite{js}). 
In this
case it is important to have the fiducial power spectrum as close to the 
actual one as possible, which again suggests iterating until this 
criterion is satisfied. We can see 
from equation (\ref{ps}) that each $T_k$ contributes 
to $P(l)$. To derive the 3-d power spectrum estimator we can 
follow the derivation of 2-d power spectrum, which leads to identical
expressions as in equations (\ref{wffll}), replacing 
$\tilde{\bi{Q}}_l$ with $\tilde{\bi{Q}}_k$, where
\begin{eqnarray}
\tilde{\bi{Q}}_k & \equiv &{\partial \tilde{S}_{\bi{ll}'} \over
\partial \Theta_l }
{\partial \Theta_l \over \partial T_k} 
=\sum_l G^2_{kl} \bi{\Pi}_l \nonumber \\
G^2_{kl}& \equiv& {\partial \Theta_l \over \partial T_k}={9 \pi\over 2}\ \Omega_m^2\ \,{g^2(\chi) \over r^2(\chi)}\
F^2(\chi,k)P^0_{\delta}(k)\ {l \over k r'},
\label{q2}
\end{eqnarray}
where $\chi$ is related to $k$ and $l$ through $r(\chi)=l/k$ and 
$r'=dr/d\chi$. Although this estimator can be applied to the 
actual measurements we obtain identical results if we apply it to 
the 2-d power spectrum estimates. 
This gives
\begin{eqnarray}
\hat{T}_k&=&{1 \over 2}F_{kk'}^{-1}\sum_l 
\bi{y}^{\dag} \bi{\Pi}_{l}\bi{y}
G^2_{k'l}
\nonumber \\
b_k&=&\sum_l b_l G^2_{kl} \nonumber \\
F_{kk'}&=&\sum_{ll'} F_{ll'} G^2_{kl}G^2_{k'l'}.
\end{eqnarray}
In the end we can put back $P^0_{\delta}(k)/\Delta \ln k$ in all the expressions 
to get an estimate of the actual power spectrum at $k$. As before we 
do not need to invert $F_{kk'}$, but can quote the convolved quantity 
$(\sum_{k'}F_{kk'})^{-1}\sum_{k'}F_{kk'}\hat{T}_{k'}$ and as long as 
the input power spectrum is close to the actual one $\hat{T}_{k'}$
will be flat and the convolved estimators will not be 
significantly biased.
 
We see that to compute the 3-d power 
spectrum we have to sum over all the 2-d power spectrum estimates, 
inverse weighted by their
covariance matrix and weighted by
their relative contribution to the given spectral bin in 3-d. This is
what one would expect, since those 2-d spectral bins
that have a large attached
error or that do not contribute significantly to the given 3-d spectral 
bin 
should be correspondingly downweighted. The method correctly includes 
both sampling and measurement errors in the analysis. Computing
the 3-d power spectrum is
no more expensive than computing the 2-d power spectrum.

\subsection{Including distance information}
So far we have assumed that the distance information is not given for 
individual objects, but only for the overall distribution of galaxies.
In this case all one can hope to measure is the projected density 
distribution and its power spectrum, as discussed in previous sections. 
If we have some additional information on the redshifts of individual 
galaxies then we can hope to extract more from the data. For example,
we may use the magnitude of the galaxy or its angular size to estimate its
distance. This can be further refined with the multicolor information,
in which case one can possibly determine the distance to each galaxy 
in the sample with better than 10\% accuracy (e.g. \cite{connolly95}).  
One obvious advantage of such information is that one can measure not
only the power spectrum, but also its evolution with time and 
this can be a strong discriminatory test of different cosmological 
models by itself (Kaiser 1992, \cite{js}). When we have this additional information 
the simple 2-d power spectrum analysis is no longer adequate, because 
different
redshift distributions of the galaxies will result in a different 
2-d power spectrum and one has to combine the information. 
In this case it is better to work with 3-d power spectrum directly,
although as we will see one finds identical results by dividing 
the galaxies into narrow bins in redshift and computing 2-d auto and cross
correlation power spectrum on these first and then 
combining them together.

Let us assume we can assign a distance and corresponding 
error estimate or a full probability distribution $W_i(\chi)$
to each galaxy individually based on their 
photometric properties in one or several bands. If we
have no information then all $W_i(\chi)$ are equal, otherwise they 
vary from galaxy to galaxy. 
The question is how to use this additional information in an optimal way. 
As before we start from the correlation matrix for the data, which
contains all the information on the statistical distribution for a
gaussian distribution with zero mean. The correlation matrix elements
are still given by equations (\ref{ps})-(\ref{corrang}) by replacing $g^2(\chi)$ 
with $g_i(\chi)g_j(\chi)$, which are obtained by integrating over the
probability distributions $W_i$, $W_j$ as in equation (\ref{shear}).
The derivative of it with respect to a given 3-d spectral bin $T_k$ is
\begin{eqnarray}
{\partial C_{ij} \over \partial T_k}&=&{\partial C_{ij} \over 
\partial \Theta_l } 
{\partial \Theta_l \over \partial T_k}=\sum_l \bi{Q}_l G^i_{kl} G^j_{kl}
=\sum_{\bi{l}}R_{\bi{l}i}G^i_{kl}G^j_{kl} R_{\bi{l}j} \equiv 
\sum_{\bi{l}}U_{k\bi{l}i}^{\dag}U_{k\bi{l}j}
\nonumber \\
G^i_{kl}&=&\Omega_m
\ \,{g_i(\chi) \over r(\chi)}\
F(\chi,k)\left[P^0_{\delta}(k)\ {9 \pi l \over 2k r'}\right]^{1/2}
\end{eqnarray}
in analogy with equation (\ref{q2}), except that we have now 
$G^i_{kl}$ depending on i-th galaxy. We introduced $U_{\bi{l}ki}=
R_{\bi{l}i}G^i_{kl}$, which can be viewed
as the analog of $R_{\bi{l}i}$ for the 2-d power 
spectrum. The 3-d estimator is now in analogy with 2-d
\begin{eqnarray}
\hat{T}_k&=&
{1 \over 2}F^{-1}_{kk'}\sum_{l'}
[\bi{U}^{\dag}_{k'\bi{l}'}\bi{C}^{-1}\bi{\kappa}]^{\dag}
[\bi{U}^{\dag}_{k'\bi{l}'}\bi{C}^{-1}\bi{\kappa}]
-b_{k'}] \nonumber \\
b_k&=&
tr[\bi{C}^{-1}\bi{U}_k\bi{C}^{-1}\bi{(N+S_b})\bi{U}_k^{\dag}]
\nonumber \\
F_{kk'}&=& {1 \over 2} 
\sum_{\bi{l}} \sum_{\bi{l}'} |\bi{U}_{k\bi{l}}^{\dag} \bi{C}^{-1} \bi{U}_{k'
\bi{l}'}|_{
\bi{ll}'}^2.
\label{ll3d}
\end{eqnarray}

The above expressions use all the distance information in an optimal way.
It is easier to compute $\bi{C}^{-1}\bi{\kappa}$ first and then multiply 
it with $\bi{G}_{kl}$, since this way the inversion only needs to be done once.
Alternatively 
one can divide the galaxies into a couple of bins in distance and 
treat all the galaxies within the bin as having the same distance 
probability 
distribution. One possibility would be to divide the galaxies in the 
magnitude bins and assign to each bin a probability distribution
$W_m(\chi)$ and corresponding $\bi{G}^m_{kl}$, where index $m$ now counts the bins and not individual galaxies.
For each bin we can compute the 2-d modes $\bi{y}_m$ using equations 
(\ref{wffll}) or (\ref{signal}). The raw 3-d power spectrum estimate
is $\sum_l (\sum_m G^m_{kl}\bi{y}_m)^{\dag}\bi{\Pi}_l (\sum_m G^m_{kl}\bi{y}_m)$
and similar expressions can be written for $b_k$ and $F_{kk}'$ as well.

As mentioned above with distance information one
may also address the question of power spectrum evolution. We can 
parametrize the power spectrum growth factor as 
$F^2(\chi)=[a(\chi)/a(\chi_s/2)]^{\alpha}$, where $\chi_s/2$ is half the 
distance to the sources,
from where the dominant contribution to the weak 
lensing signal is coming (\cite{js}). This form minimizes the degeneracy 
between the amplitude of the power spectrum and the growth factor. We can 
ask how to extract $\alpha$ from the data. Expanding the likelihood
function around the maximum gives the estimator of equations 
(\ref{fisher})-(\ref{wffll}), 
replacing $\bi{Q}_l$ with $\bi{Q}_{\alpha}$, the 
derivative of the correlation function with respect to $\alpha$, 
$\bi{Q}_{\alpha}=\partial \bi{C} / \partial \alpha= \bi{C}_{,\alpha}$, where 
$\bi{C}_{,\alpha}$ is the correlation function computed by using $F^2(\chi) \ln [a/
a(\chi_s/2)]$ instead of $F^2(\chi)$. 
For a narrow dynamic
range in distance and scale 
the growth factor will be degenerate with the shape of 
the power spectrum, which will be reflected by the covariance terms 
$F_{l\alpha}$ and one has to invert the matrix to obtain an estimate 
of the error on the growth factor that does not depend on the power 
spectrum. 

\section{Power Spectrum Estimation: simulated data}
In this section we apply the formalism presented in previous section to 
estimate the power spectrum from simulated 
weak lensing data. To generate a simulated section of a field
we first make a random realization of the convergence 
$\tilde{\bi{\kappa}}$ in 
Fourier space and compute the two shear components based on equation 
(\ref{shearkappa}). We then Fourier transform the shear to real space and select 
a smaller area as our observed field (we pay special attention not to
select the area to be an integer fraction of the total simulated area
to avoid any effects from periodic boundary conditions). We randomly 
generate the positions of galaxies with a specified surface density 
and randomly generate noise for each component of the shear from 
a gaussian distribution with a specified rms error. 
We add this value to the value of the shear at the
galaxy position and use this map as our simulated data. We use 
$\Gamma=\Omega h=0.25$ and $\sigma_8=0.6$ as our fiducial cosmological 
model and place all the galaxies at $z=1$. These numbers should be 
typical for a several hour exposure at a 4m class telescope with a
limiting magnitude around $I=26$. We then compute the power spectrum 
estimator using the expressions in previous section using the 
expressions in Fourier space, because they require less computational 
time than the corresponding expressions in real space. 

\begin{figure}[h]
\vspace*{8.3 cm}
\caption{2-d power spectrum reconstruction of filled 1 square degree field
(left) and $100$ square degree 
sparsely sampled field (right). The points with 
error bars connected with the line are the estimators, the thick solid
line is the input power spectrum, thin solid line is the corresponding
linear power spectrum and dashed line is the reconstructed power spectrum
of the $45^{\circ}$ rotated data.  }
\includegraphics{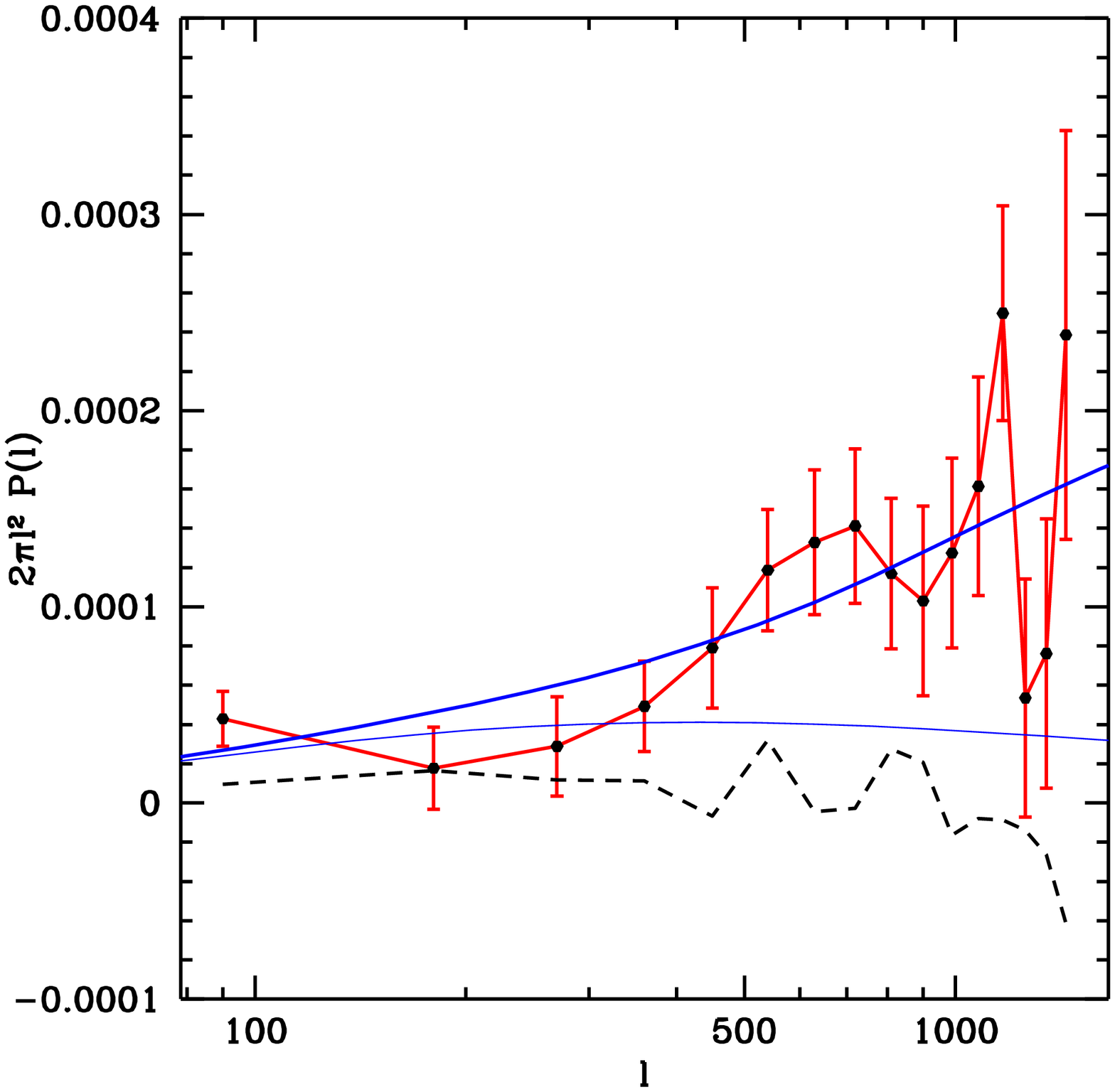}
\includegraphics{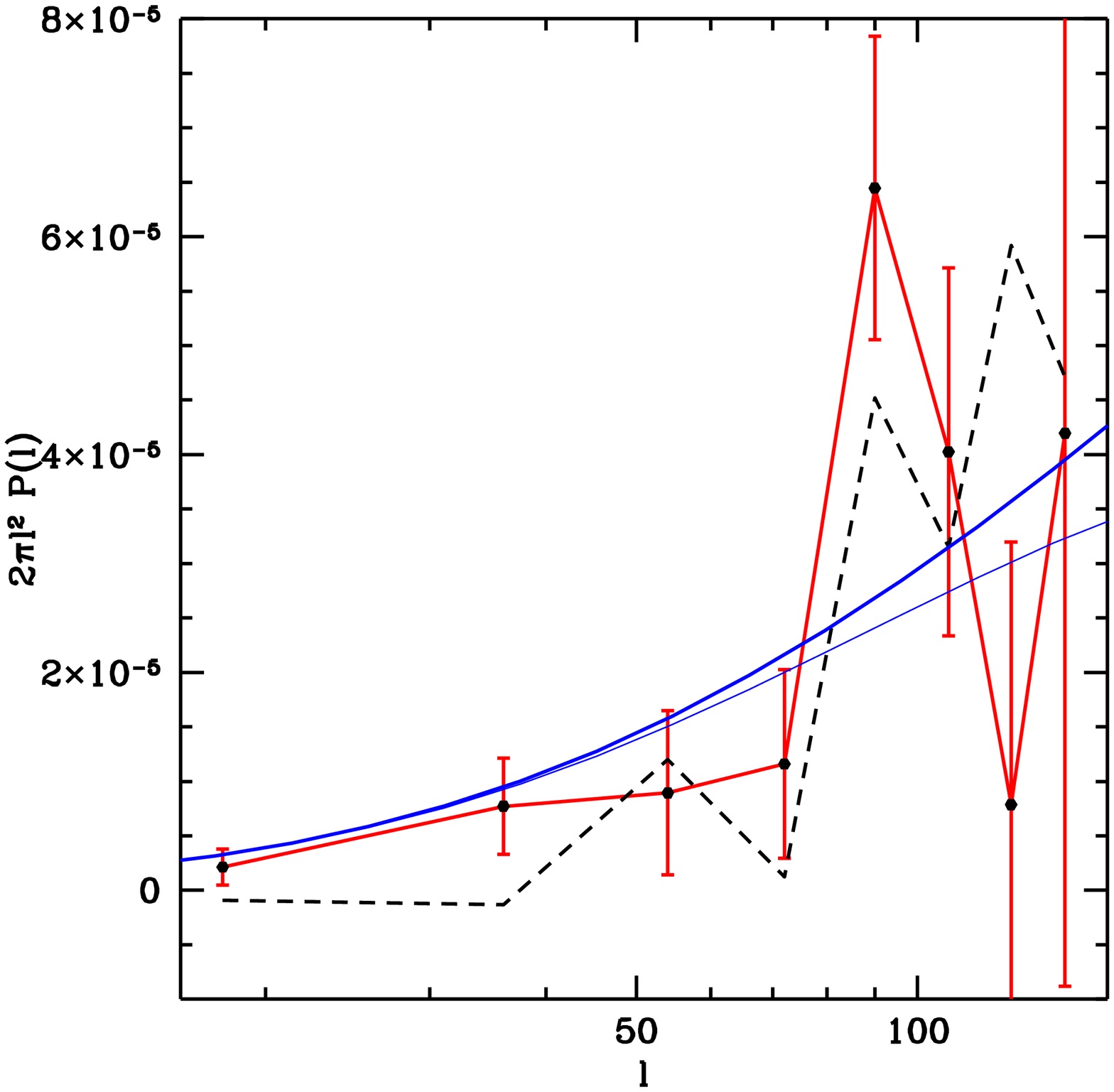}
\label{fig4_1}
\end{figure}

Figure \ref{fig4_1}a (left)
shows the power spectrum measured on a 1 square degree
field, using $2\times 10^5$ galaxies with rms ellipticity
error of $\langle e_i^2 \rangle^{1/2}=0.4$. 
The estimators are 
given as points with attached error bars computed from the Fisher matrix,
connected with a solid line.
We compared the errors computed analytically with the errors obtained 
from Monte Carlo realizations and found excellent agreement. The errors
on small scales are dominated by noise, while on large scales the 
dominant contribution comes from sampling (cosmic) variance. 
These errors are based
on gaussian approximation, which underestimates them in the nonlinear 
regime. To estimate the scale where this may be important
we plot on the figure both the input nonlinear 
power spectrum (thick solid line) and the linear power spectrum
(thin solid line). We see that on the largest scales the two agree and
the sampling variance is correctly estimated, while on smaller scales
the nonlinear power spectrum significantly deviates from the linear 
one and the sampling variance would be underestimated. However, where 
this becomes important noise will often be
the dominant source of error. Noise 
can be well approximated with the gaussian probability distribution 
because of the large 
number of galaxies, even if the intrinsic distribution of ellipticities
is not gaussian. For the example we 
have chosen uniform dense sampling of galaxies 
prevents any significant amount of aliasing of small scale power and 
we find almost identical results ignoring aliasing terms in 
equations (\ref{wffll}). This is no longer the case when the sampling is sparse,
as shown below.

We can also test whether the signal is gravitational by rotating the 
galaxies by $45^{\circ}$.
The dashed curve in figure \ref{fig4_1} 
is the result of power spectrum estimation 
on the data rotated by $45^{\circ}$. 
For perfect data this rotation would eliminate the 
gravity mode and only excite the vorticity mode (\cite{stebbins96}). 
Since there were no
vorticity modes put in the simulation (and indeed almost none are expected
to be present in the real data) one can hope to detect any systematic
problems through this test. The dashed curve in figure \ref{fig4_1}
is the result of power spectrum estimation
on this rotated data. 
The resulting power spectrum is significantly lower than the 
one using original data, which confirms the usefulness of this test. 
Note however that the power spectrum 
is not consistent with zero and there is an 
excess of power present in this mode. The reason is aliasing of power 
from the 
gravity mode to the vorticity mode. If this aliasing is not included in 
the analysis (as it was not in this example) it will artificially create
power in this mode. We tried to analyze the data without putting the power
into the gravity mode and found 
the recovered power spectrum consistent
with zero. If however there is true power present in the data it will 
act as an additional noise for the
vorticity mode and has to be taken into account. 
This example shows that one has to be careful
to use the rotation test as a way to monitor the contamination of the data 
by some source that adds equal power to the two modes. One has to account
for the aliasing first by computing the power spectrum of the gravity 
mode and use it to compute the total noise contribution in analogy 
with the bias term in equation (\ref{wffll})
\begin{equation}
b_{45^{\circ},k}=
tr[\bi{S}^{-1}\bi{D}^{-1}\bi{R}_{45^{\circ}}^{\dag}\bi{N}^{-1}\bi{C}
\bi{N}^{-1}\bi{R}_{45^{\circ}} \bi{D}^{-1}\bi{S}^{-1}].
\label{vorb}
\end{equation}
Here $\bi{C}$ is the correlation matrix (equation \ref{corr}) and
$\bi{R}_{45^{\circ}}$ the response matrix for vorticity modes, which can
be obtained from $\bi{R}$ by replacing 
$\{\cos 2\phi,\sin 2\phi\}$ with $\{-\sin 2\phi,\cos 2\phi\}$. 
Similarly, Fisher matrix also has to be modified to account for this 
additional noise term.

On large scales the errors are completely dominated by sampling variance
and as suggested by Kaiser (1997) sparse sampling would be very useful
to reduce the errors. As an example we computed the power spectrum by 
randomly placing 100 observations over an area of 100 square degrees 
(figure \ref{fig4_1}b). In this example
each observation consists of a single exposure detecting 10000 galaxies,
which can be grouped together into a single measurement with rms noise
$\langle e_i^2 \rangle^{1/2}=0.004$. An example would be a number of
short (a few minute) exposures on a $(30')^2$ field, with a sampling 
factor of 4, or a longer exposure with more galaxies per exposure
and correspondingly lower sampling rate, both of 
which would only require a couple of nights with the 
composite CCD cameras. This choice of sampling is not necessary the
optimal one for measuring the power spectrum, which in detail
depends on the properties of 
noise, signal and aliasing power (\cite{Kaiser97}). 
Probably even sparser sampling would give an even higher reward. Even so 
it is clear that this strategy offers a significant promise to measure
the power spectrum on very large scales where the potential rewards are
largest. 

When the data are sparsely sampled it is important to properly remove the
small scale aliasing: 
the power spectrum estimator
without the subtraction of the aliasing term would be strongly biased 
towards the large values and since this term is only known to the extent 
that it has been measured well on small scales 
it can easily dominate the errors attached
to the estimator. Similarly, the Fisher matrix had to be inverted to 
compare the estimators to the input (rather than convolved) power spectrum,
otherwise
the estimators would have been biased (this effect can be minimized by 
estimating not the power spectrum directly, but only its departure 
from some fiducial power spectrum, as discussed in section 2.4). 
This is again a consequence of
strong mixing between the modes in different power spectrum bins.
Note that for such sparse sampling $45^{\circ}$ rotation test gives 
a lot of power in the vorticity mode, specially 
on small scales.  
For even sparser sampling the vorticity mode easily exceeds
the input power spectrum because of the small scale aliasing.  
This has to be taken into account for this test to be a useful diagnostic
of any nongravitational contribution. One can do this 
by using equation (\ref{vorb}) 
when estimating the noise.

\begin{figure}[h]
\vspace*{8.3 cm}
\caption{3-d power spectrum reconstruction of filled 1 square degree field
(left) and $100$ sparsely sampled field (right). The points with 
error bars connected with the line are the estimators, the thick solid
line is the input power spectrum. }
\includegraphics{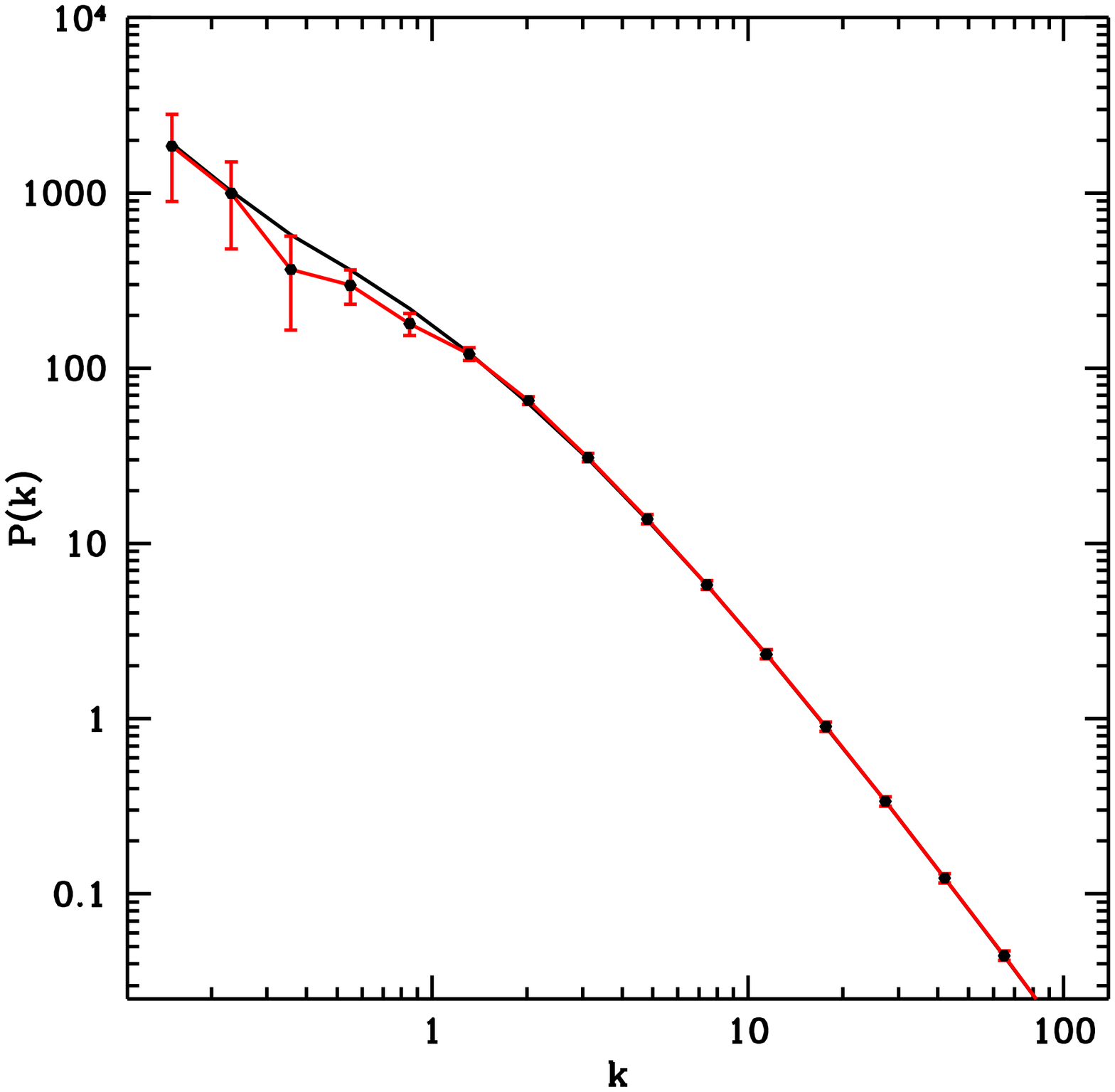}
\includegraphics{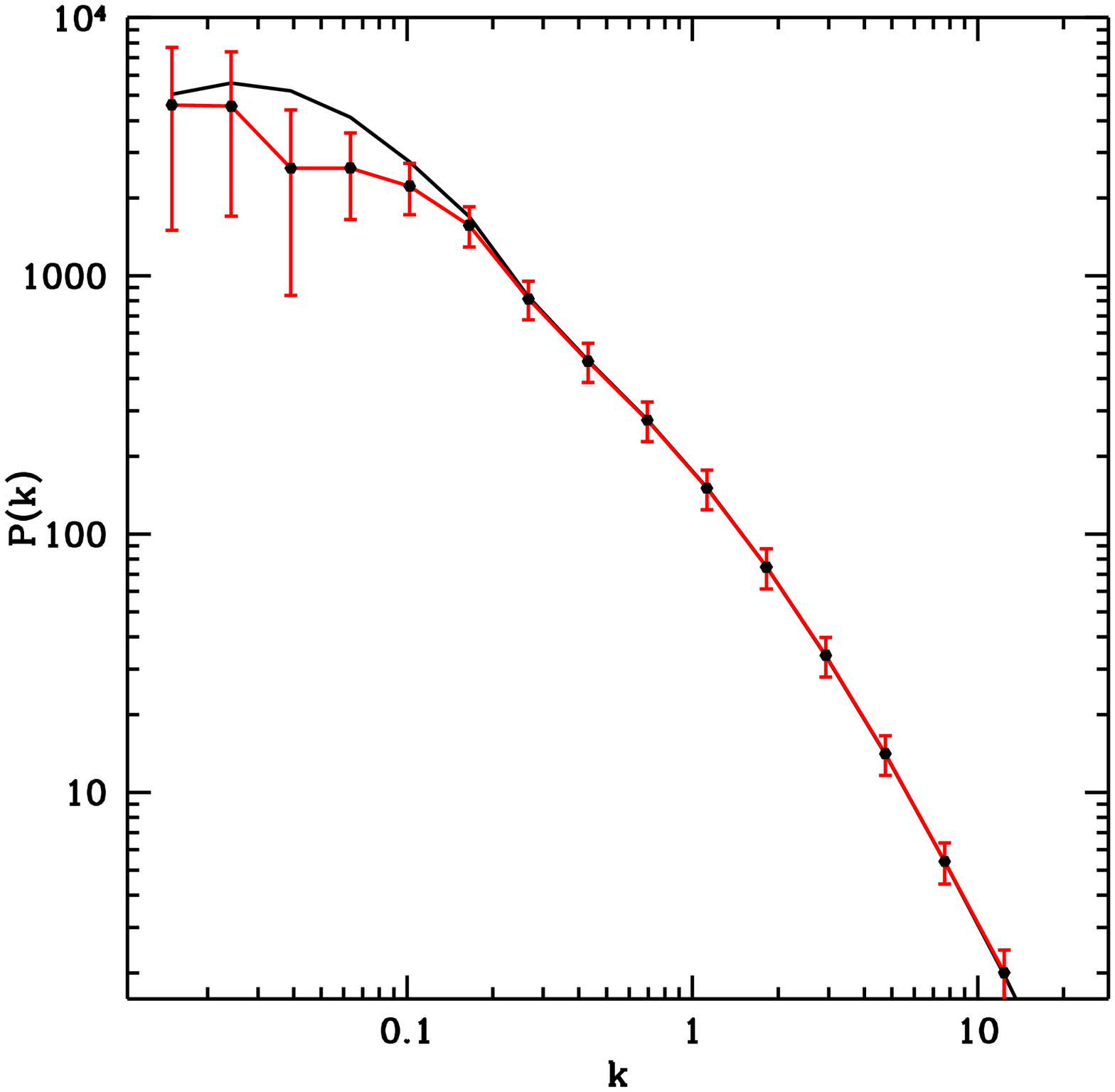}
\label{fig4_2}
\end{figure}

We also tried the 3-d power spectrum estimation using the approach 
outlined in section 2.3. The results for the 1 square degree filled
and 100 square degree sparse survey 
corresponding to examples in figure \ref{fig4_1}
are shown in figure \ref{fig4_2}. We show 
the results for the convolved power spectrum and individual
estimates are strongly correlated. This, logarithmic binning and  
the use of logarithmic scale make the results visually more impressive 
than in figure \ref{fig4_1}b, although the information content is 
the same. Nevertheless, the results are quite impressive and 
an even larger (and sparser) survey would give an even bigger
reward of determining the turnover in the power spectrum quite 
precisely. We verified the analytic error estimates from the Fisher 
matrix by comparing them to Monte Carlo estimates 
and they agree quite well. On small scales where the nonlinear 
effects become important the sampling error is 
underestimated because we used the
gaussian approximation and one has to 
correct for that if necessary.
\section{Mass density reconstruction: simulated data}
In this section we apply the formalism developed in section 2
to reconstruct the projected dark matter distribution from weak lensing 
observations. We start with the reconstruction of large scale structure,
where the techniques developed here work best. We then apply the same 
methods to the cluster reconstruction, pointing out its advantages and 
disadvantages and comparing them to other reconstructions. 
For this case we also 
generalize them to include nonlinear effects and discuss how to treat 
the redshift distribution of the sources. We will assume throughout 
this section that the power spectrum of the data is known in advance,
using the techniques developed in section 3. The power spectrum is only
known where the signal exceeds the noise, beyond that one has to assume
its form by 
parametrizing it with a power law. 
We will use in the applications the actual power spectrum
obtained from the simulated data directly. This of course cannot be 
obtained with the real data, but it turns out that in practice the reconstruction 
is not very sensitive to this, because the spectral index of the data
power spectrum differs significantly from the noise spectral 
index 0 and so the only important parameter is the scale 
where the signal power spectrum exceeds the noise. Because we are 
expanding the data with modes that have periodic boundary conditions
in a box it is better to use a somewhat larger box than the size of 
the observed field and zero pad the rest. 
For the examples here we used 10\% padding on each side and in most
cases this eliminated the spurious effects associated with 
periodic boundary conditions. If this is not sufficient one can always 
increase the size of zero padded area.

\subsection{Large Scale Structure Reconstruction}

\begin{figure}[p]
\vspace*{17.3 cm}
\caption{Reconstruction of a random portion of the sky. 
The sidelength is 1 degree and $2 \times 10^5$
randomly generated galaxies have been used for reconstruction.
Top left are input data, top right smoothed
input data, bottom left WF reconstruction and bottom right  
$45^{\circ}$ rotation test. 
Most reconstructed structures have surface density of a few percent.}
\includegraphics{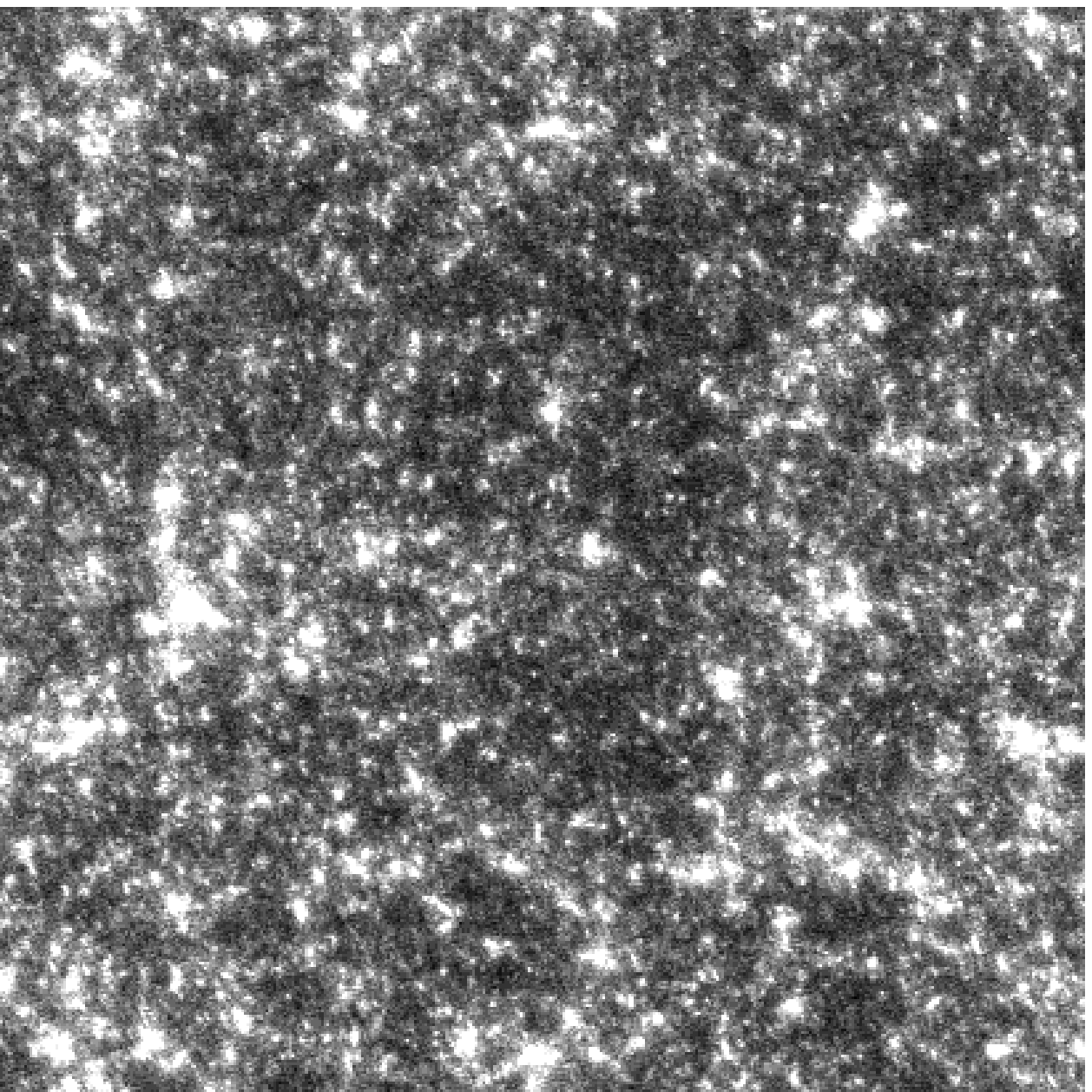}
\includegraphics{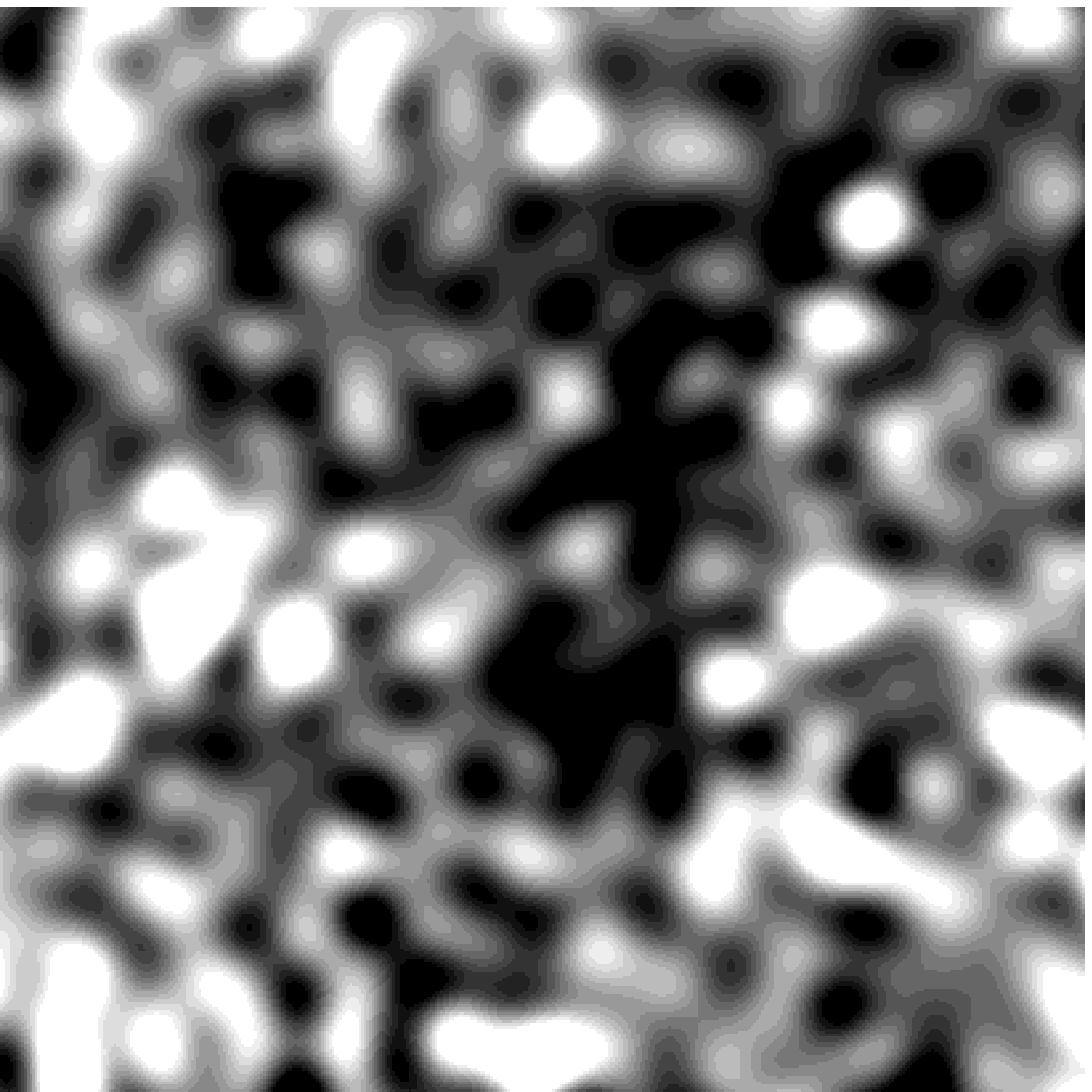}
\includegraphics{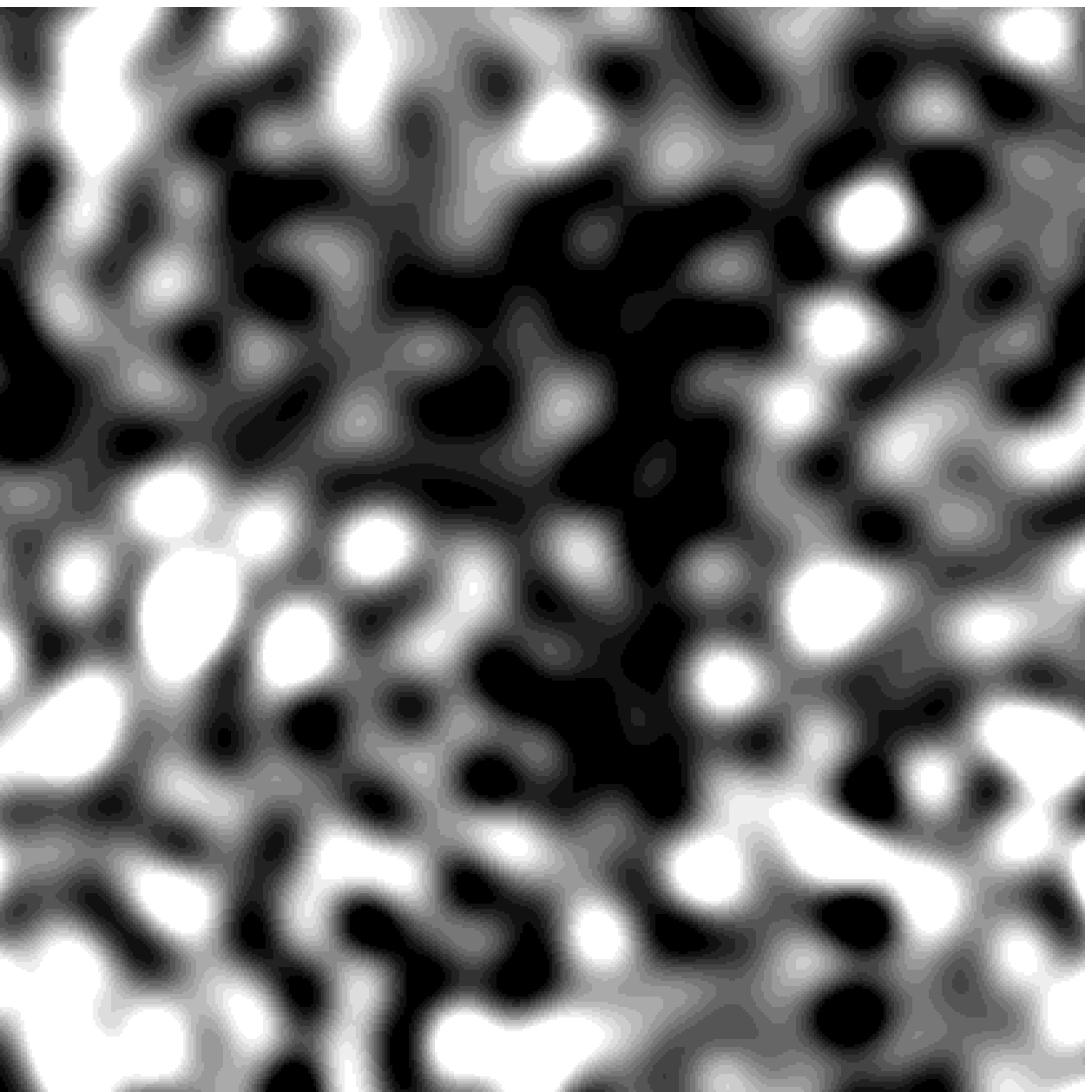}
\includegraphics{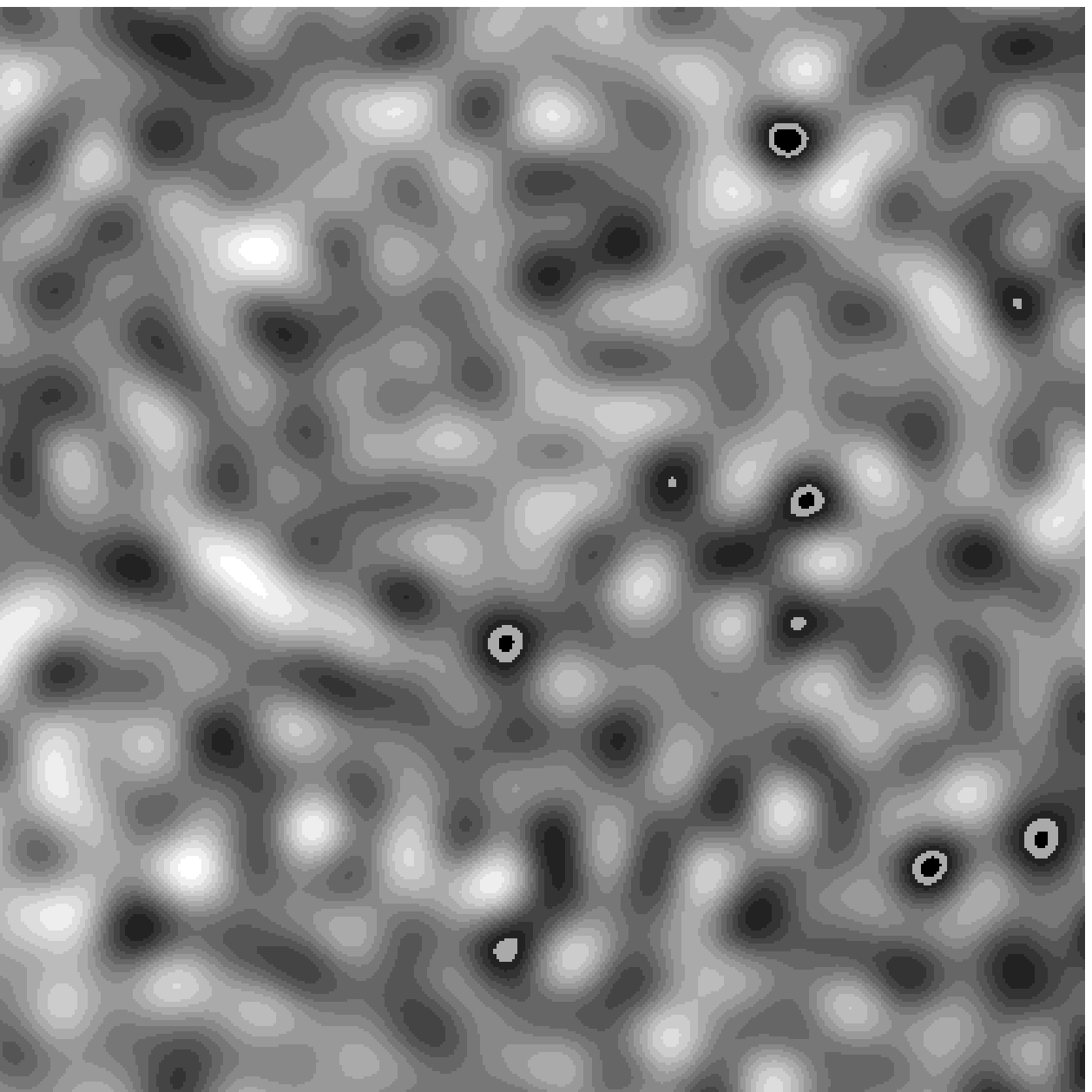}

\label{fig3_4}
\end{figure}

We begin with the reconstruction of large scale structure. For 
this purpose we generated a projected density map from an N-body
simulation. 
The simulated data in figure \ref{fig3_4} were obtained by 
randomly placing $2\times 10^5$ galaxies on a 
1 square degree field. 
Upper left panel of figure \ref{fig3_4} 
shows the input data, which have a lot of small
scale structure. Smoothing the input data  at the scale where 
signal power spectrum equals the noise power spectrum produces upper right 
panel of figure \ref{fig3_4}. WF reconstruction is given in 
lower left panel of figure \ref{fig3_4}. We see that WF strikes
the balance between the resolution and signal to noise ratio. The 
reconstruction is heavily smoothed because of the low signal 
to noise on small scales. On the other hand, most  
of the reconstructed
structures are real (compare upper right with lower left panel), 
because WF preferentially keeps the modes that are 
above the noise. WF is thus particularly useful for identifying
the large scale structures, such as filaments or superclusters in 
the data. Rotation of the galaxies by $45^{\circ}$
eliminates most of the signal and the reconstructed field
is now significantly lower than  
the typical structures in the reconstruction, as given in lower 
right panel of figure \ref{fig3_4}. This does not however
completely eliminate the signal, because as discussed in section 3 
aliasing of power from the gravity mode into the vorticity mode
makes the latter nonvanishing even in the limit of small noise. 
The effect is rather small because of the 
large number of background galaxies, but would be more important
if sampling was sparse.

\subsection{Cluster Reconstruction}
The methods we developed in this paper 
are optimal for gaussian random fields, but 
they can also be applied to nongaussian situations, such as cluster
reconstruction from weak lensing. In this case WF by definition still
minimizes the variance in the class of linear estimators as shown in 
section 2.1 and so remains a useful reconstruction technique. 
To test it we apply it to reconstruct a massive cluster
obtained from an N-body simulation (figure \ref{fig3_1}a). We have rescaled 
the projected mass density of the cluster in units of critical density, 
so that it is nearly critical. The cluster core is 
very concentrated and shows double nucleus structure.
The comoving size of the box is 3.2$h^{-1}$Mpc and we will take the
area to be 10' across with 5000 galaxies in it. 
In this subsection we want to compare the reconstruction properties of
various filtering methods and we will ignore
the nonlinear corrections. Those will
be addressed in the next subsection.

\begin{figure}[t]
\vspace*{7.3 cm}
\caption{Power spectrum of signal (thick solid curve) and noise 
(thick dashed curve), obtained by computing the power spectrum 
with only cluster or only noise, respectively.
Only for long wavelength modes signal exceeds the noise. 
Thin lines give the power spectrum of residuals,
(the difference between the true and reconstructed modes)
for WF (solid) and white noise prior WF (dashed). 
The two have similar 
noise properties for long wavelength modes, while for short wavelengths
white noise prior gives much larger noise variance and reduces to the 
noise power spectrum, because the true modes are small. Conversely for
WF the power spectrum of residuals reduces to the signal power spectrum 
on small scales because the 
reconstructed modes are small. 
}
\includegraphics{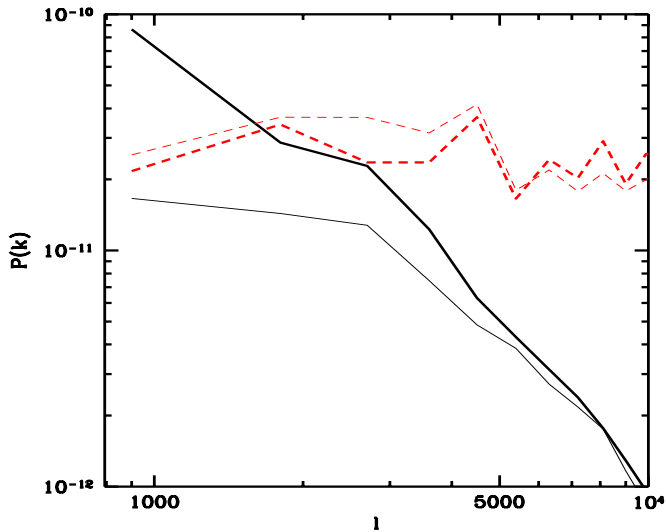}
\label{fig3_3}
\end{figure}

\begin{figure}[p]
\vspace*{17.3 cm}
\caption{The dimensionless surface mass density $\kappa$ for the 
simulated cluster. The sidelength is 10' and 5000 
randomly generated galaxies have been used for reconstruction.
The contour levels are 0.1,0.2,0.3,0.4 (solid) and 0.01,0.03,0.05,0.07 
(dashed). From top left to bottom right the plots are: input data, WF reconstruction,
WF reconstruction with uniform weighting approximation, $45^{\circ}$ rotation test,   
white noise prior WF reconstruction and WF nonlinear reconstruction. WF
heavily smooths the data, because the cluster is only a small part of
the observed region and does not add a lot of power to the power 
spectrum.}
\includegraphics{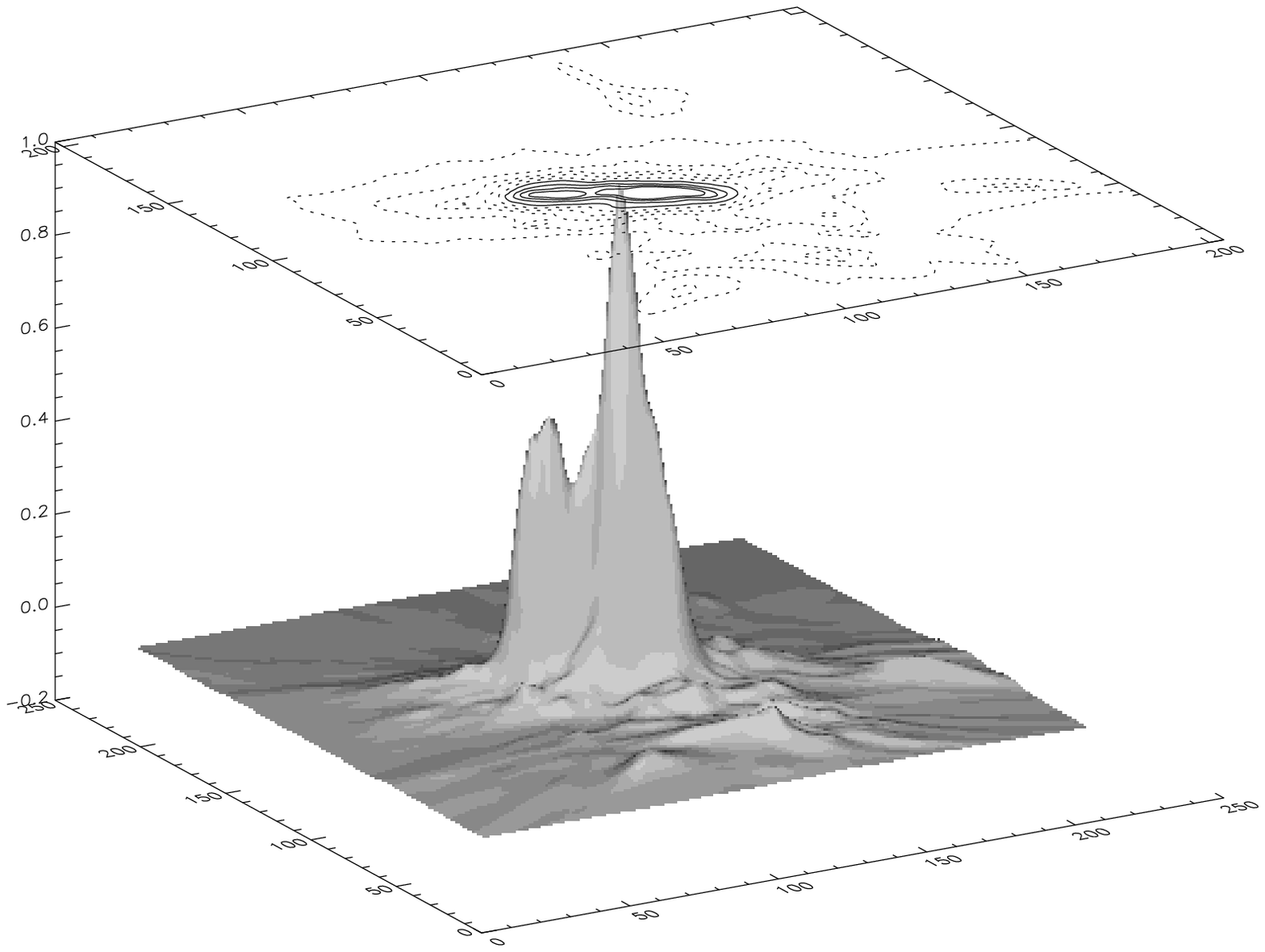}
\includegraphics{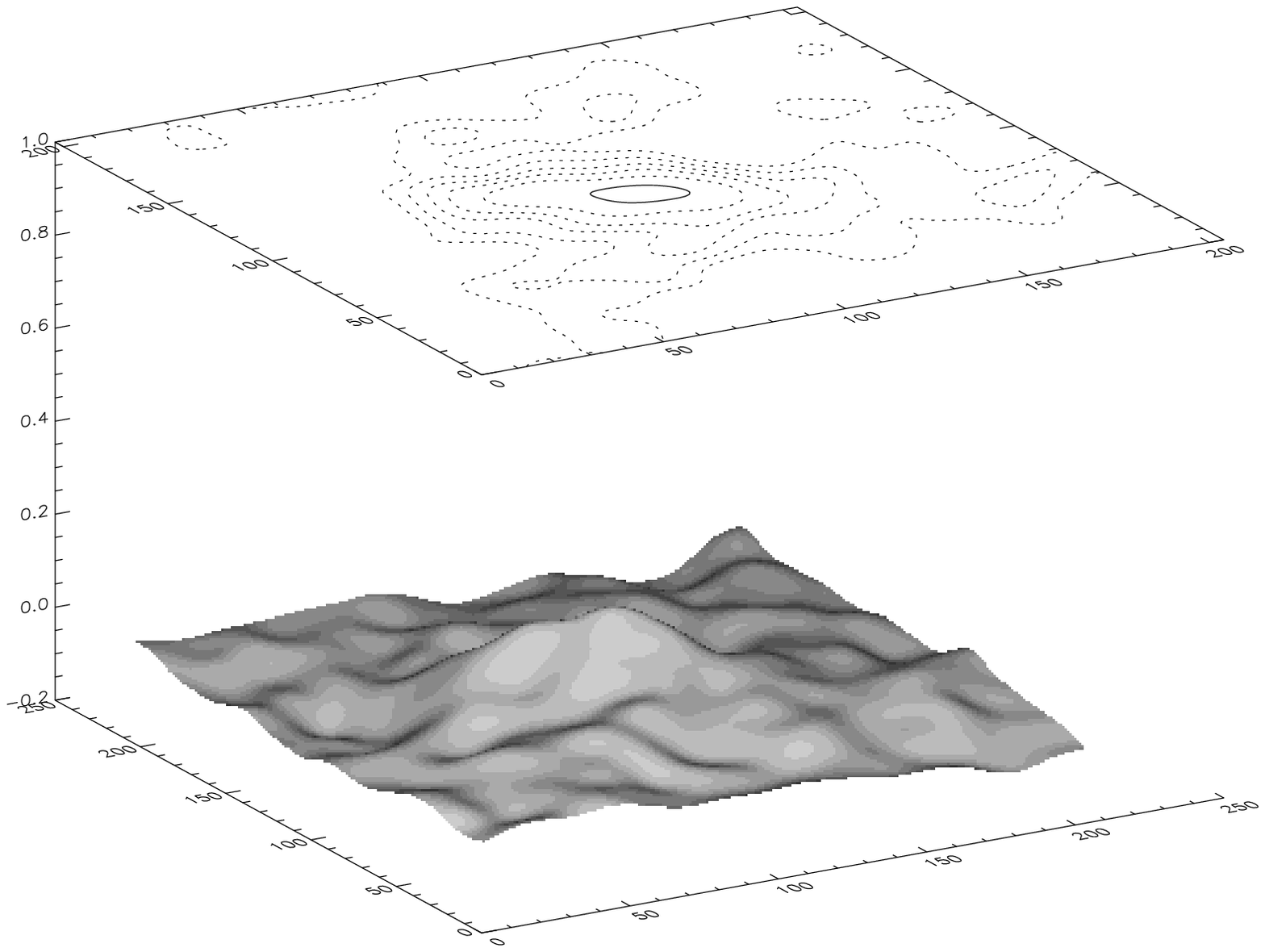}
\includegraphics{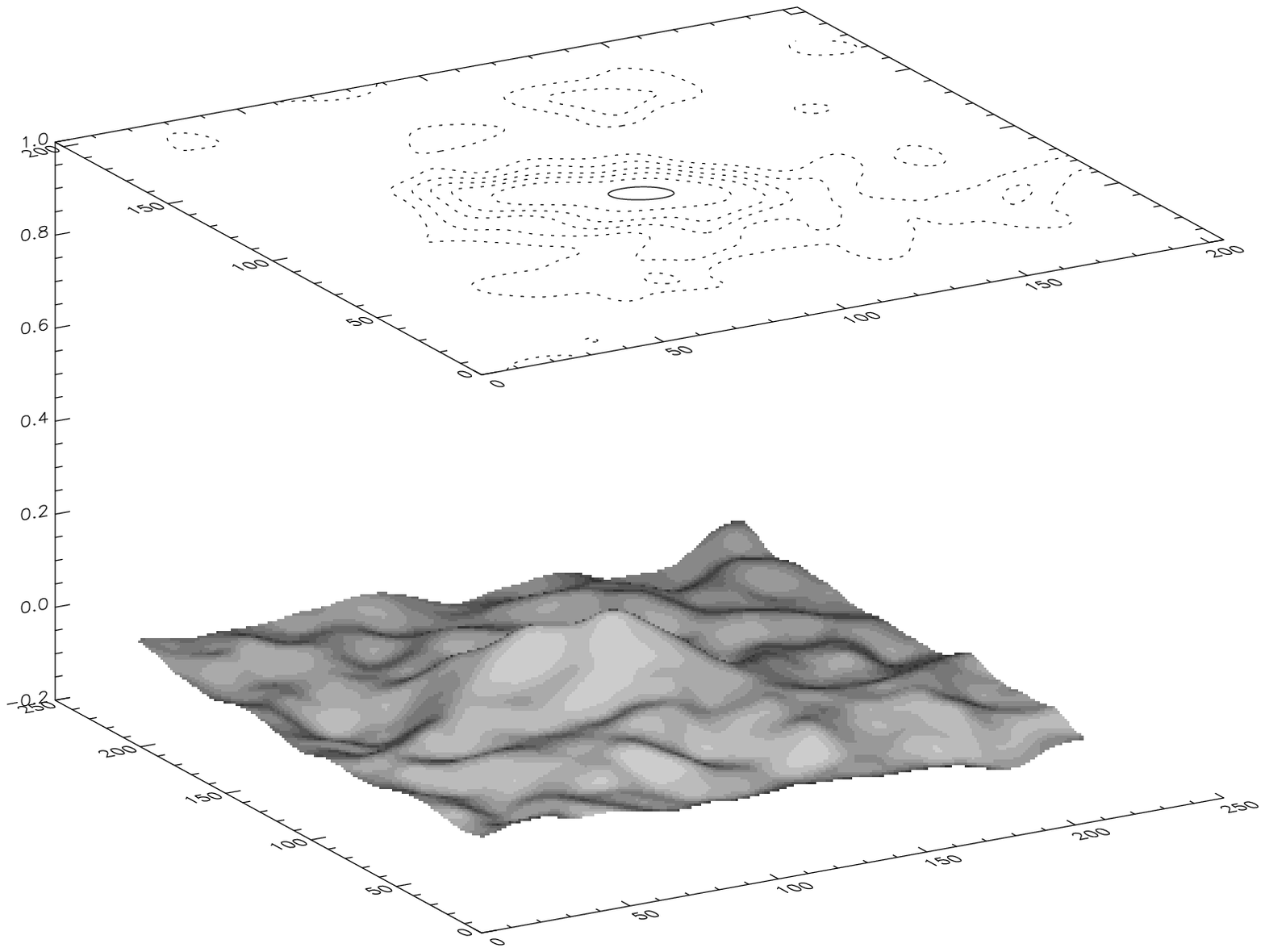}
\includegraphics{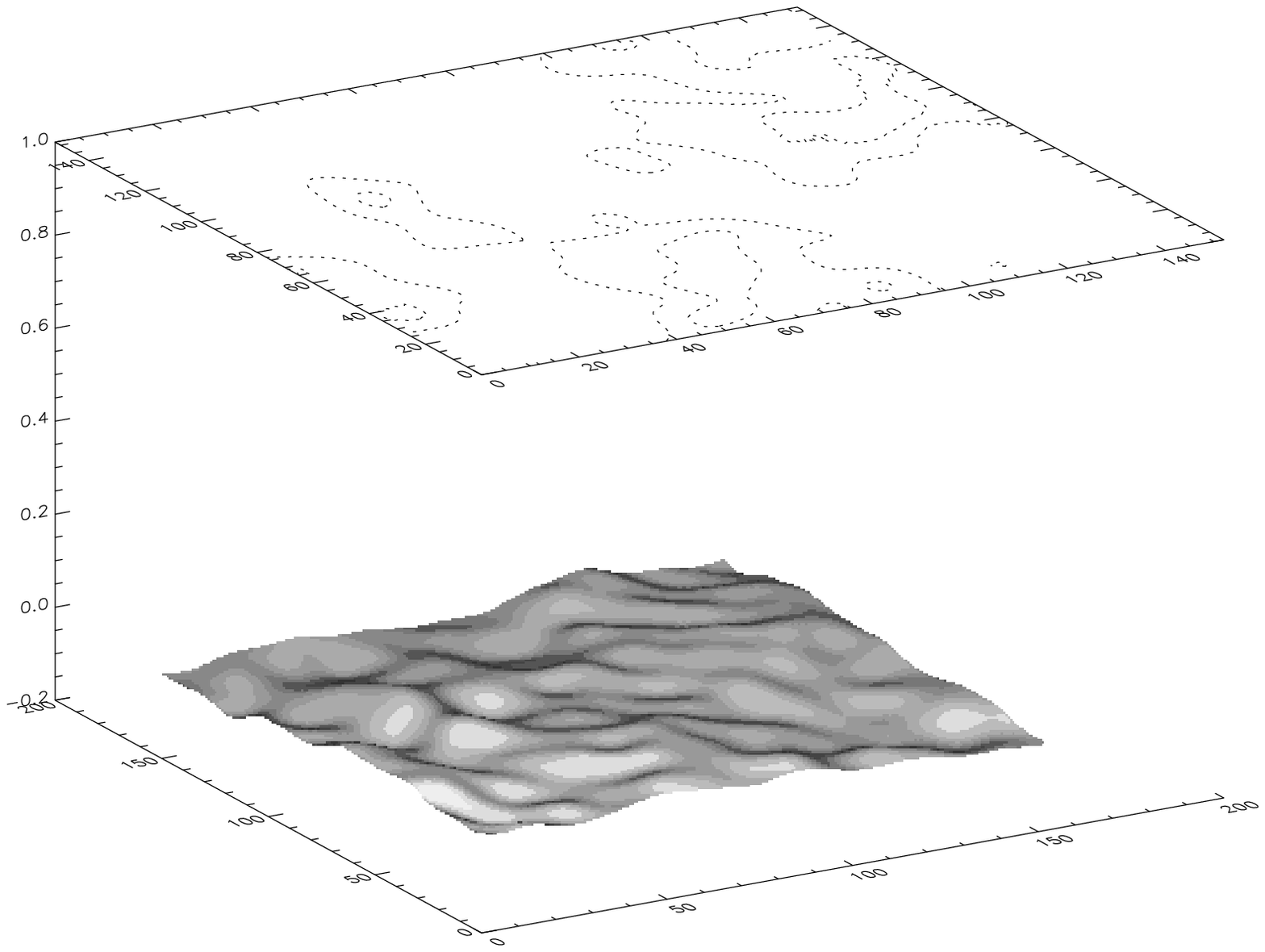}
\includegraphics{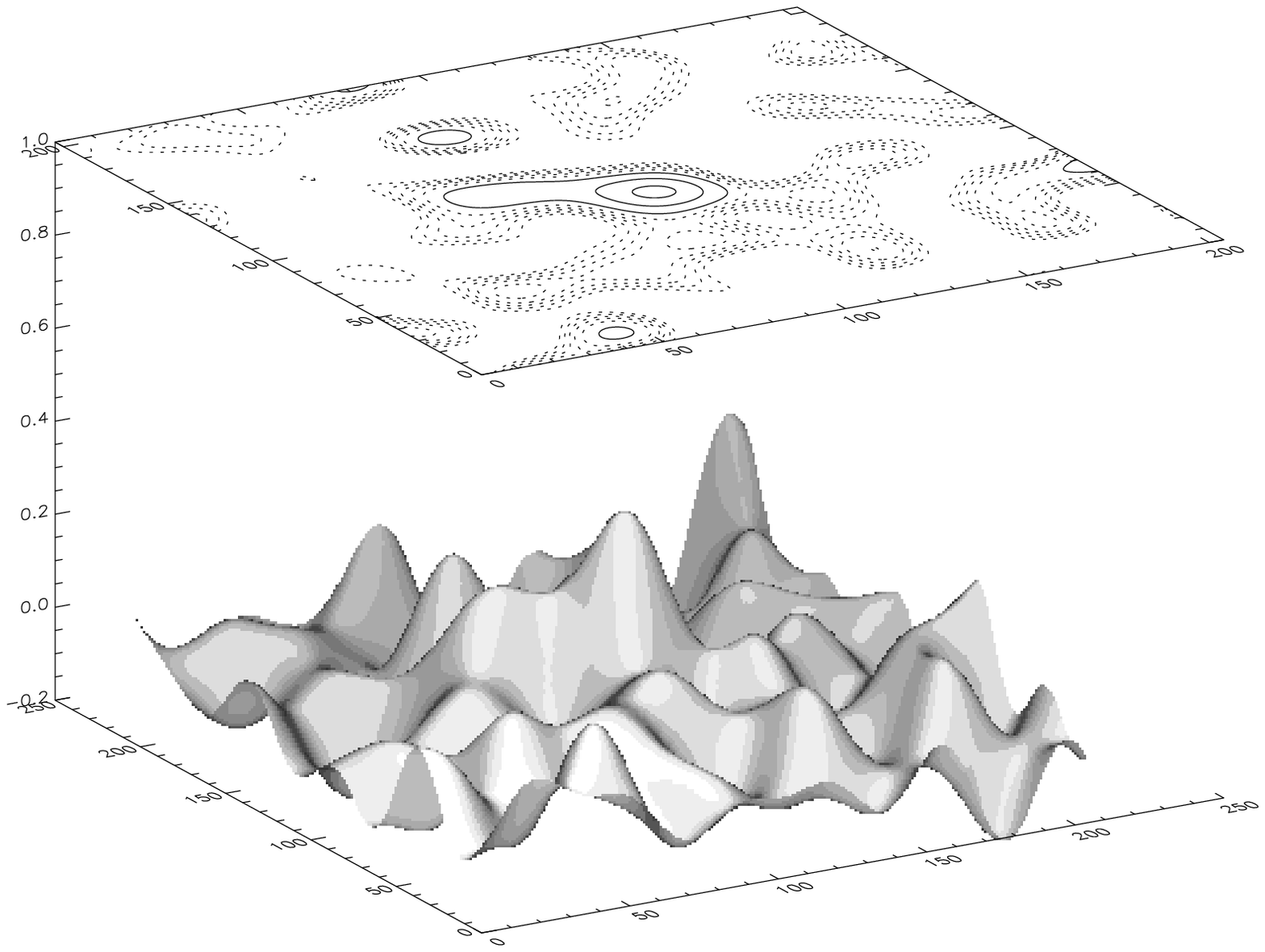}
\includegraphics{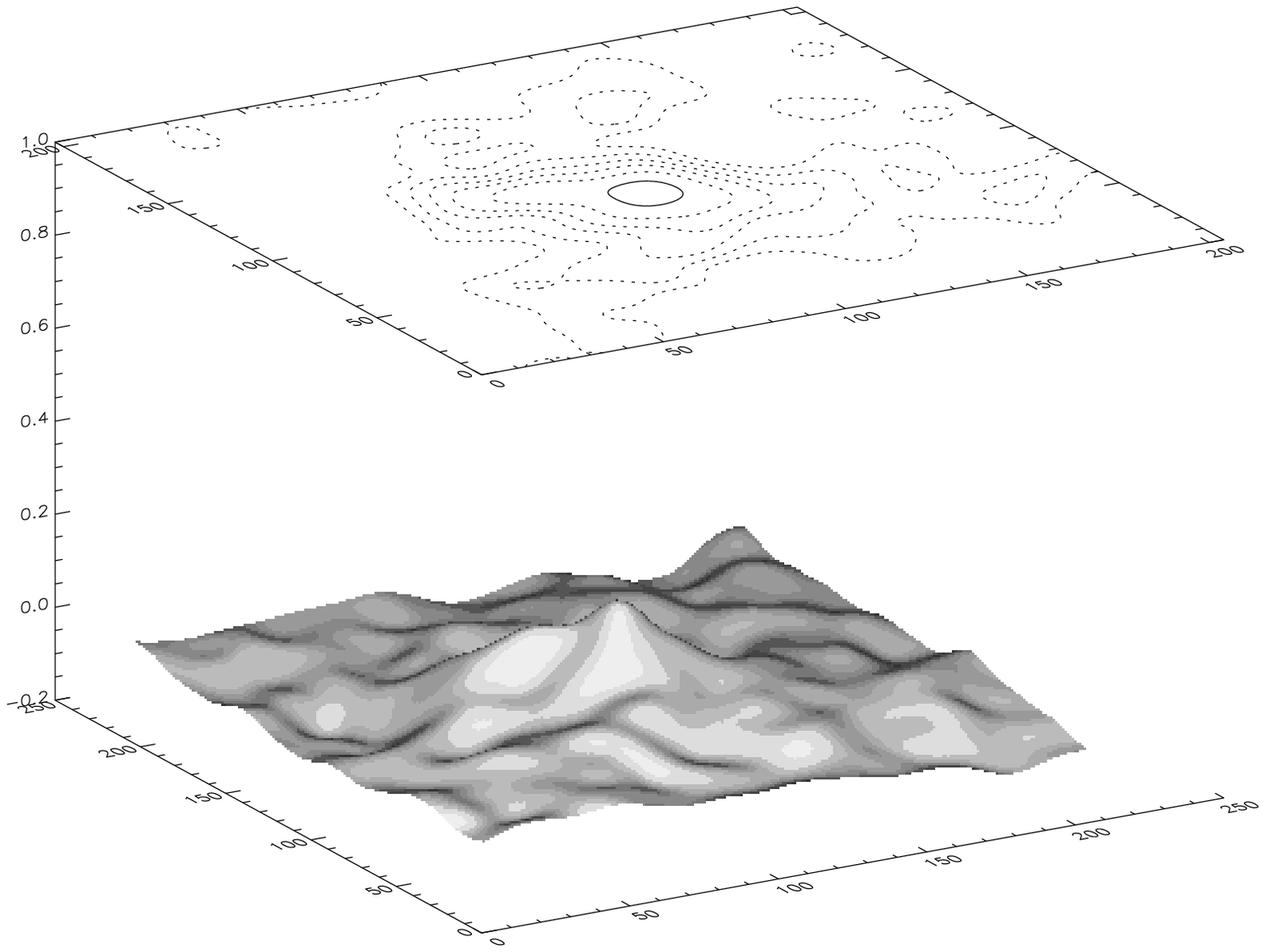}

\label{fig3_1}
\end{figure}

There exist a number of cluster 
reconstruction techniques in the literature.
Squires \& Kaiser (1996)
present a comprehensive review of most of these and conclude
that their so-called 
maximum probability model is the best in the sense that
it has the least amount of long-wavelength noise in the noise
power spectrum.
This is not surprising, since their model is 
identical to WF (in Fourier space), except that they advocate the 
prior power spectrum to be white noise with adjustable amplitude. The
latter was taken to be  
larger than the noise, so that it does
not suppress the modes on any scale 
(the only reason \cite{sk} add the theoretical
prior is to regularize the inversion of matrix $\bi{D}$).
Because for long wavelength modes the actual signal is typically
larger than the noise (as shown in figure \ref{fig3_3} for the cluster
used here) the results will be similar to WF and
both methods will reconstruct long wavelength modes
without any suppression.
However, on scales where the power spectrum drops below the noise the 
two methods differ: white noise prior does not
filter the modes and is reconstructing mainly noise, while WF
suppresses the modes, returning zero in the regions of low
signal to noise. This is what one expects from a minimum variance 
method, since zero deviates from the true field less than the random 
noise does when the power spectrum of the latter exceeds the former. 
This is shown in figure \ref{fig3_3} for the simulated 
cluster. For short wavelengths white noise prior
is no longer optimal in the sense of minimizing the variance and 
so in this sense WF is a minimum variance reconstruction among 
linear estimators even for a cluster. But as we show next this 
may not be the most desirable feature in the reconstruction and 
other methods (such as the one 
investigated by Seitz and Schneider 1996) may result in a better 
overall reconstruction.

Because the power spectrum has less power on small 
scales than the white noise one expects WF to result in 
a smoother reconstruction than the one assuming white noise as a
theoretical prior.
WF reconstruction of the cluster is shown in upper right panel of 
figure \ref{fig3_1} and indeed
is rather smooth compared to the white noise prior WF reconstruction.
Note that the uniform weighting with variable theoretical variance 
(equation \ref{wf}) shown in the middle
left panel of figure \ref{fig3_1} gives almost identical results, 
but is much easier to compute.
WF reconstruction 
shows a clear signature of the cluster above the background or above 
the $45^{\circ}$ rotation reconstruction in the middle right panel of 
figure \ref{fig3_1}, which does not show any evidence of a cluster.
The noise is mostly eliminated from the map and most of the large 
structures one sees are real.
These properties make WF to be a useful method even
for cluster reconstruction. However, there is a high price 
to pay for this.
By suppressing the small scale modes and so the 
noise across the field WF also suppresses 
the central peak of the cluster. 
Some amount of smoothing is of course inevitable if
the data do not have sufficient resolution, but WF tends to heavily
smooth the data by suppressing all the small scale modes,
regardless of their position.
The nongaussian nature of the cluster center may give sufficient
signal to noise to reconstruct it, but because all of the power
on small scales is concentrated in this small region it may not give
statistically significant excess above the noise in the power spectrum
analysis. This in fact becomes exacerbated as we increase the area of
the reconstruction, because by attempting to reconstruct the whole 
area (with no power on small scales 
over most of the region) WF will further suppress 
the central part of the cluster. For some applications one may be 
interested more in the central part of the cluster and willing 
to accept higher levels of noise at the outskirts of the cluster
in which case WF will not be ideal method.

For cluster reconstruction it is therefore worth extending 
the family of linear estimators to make the prior 
power spectrum a free parameter. One example in this class is
white noise
prior advocated by Squires \& Kaiser (1996), which can reconstruct 
the structure on small scales.
It should be stressed that with white noise prior one still has to impose a 
high frequency cutoff, otherwise the noise on small
scales completely dominates the reconstruction. 
The way Squires \& Kaiser (1996) impose this is by restricting themselves
to a small number of modes in the expansion, although the actual number was 
left unspecified. 
This method therefore amounts to a simple low-pass
filtering, where the filter is a step function, as opposed to the WF,
where the filter is more gradual and depends on the actual signal to noise 
ratio in the power spectrum of the data. If we extend the filtering scale
beyond the scale where signal and noise power spectra are equal we obtain 
the reconstruction shown 
in lower right panel of figure \ref{fig3_1} for one particular choice 
of cutoff in $k$. 
Most of the reconstructed structure on small scale is noise,
which clearly
does not minimize the variance as defined in equation (\ref{res}) and
so is not optimal in this sense. On the other hand, the central peak of 
the cluster is now significantly higher and its detection is more
significant than in the case of WF. 
In this case reconstruction with smaller filtering scale 
may be more acceptable, despite being significantly noisier. 
Nonlinear methods such as maximum entropy method (\cite{nar}) 
are likely to do even better in such applications.
Note that the spike at the corner of the white noise prior reconstruction 
is an artifact 
of the periodic boundary conditions and would be eliminated if one used
20\% zero padding on each side of the box. This example shows that 
one has to be careful about the size of the zero padding area,
which will be somewhat dependent on the type of the filter one is using.

\begin{figure}[t]
\vspace*{7.3 cm}
\caption{Same cluster as in figure 3.1 assuming it is twice
larger in the sky.  
The sidelength, number of galaxies and contour levels are the same as in 
figure 3.1. Left is the input surface density, right the 
nonlinear WF reconstruction. In this case WF smooths the data less than 
in figure 3.1, because the cluster is a dominant part of the 
observed region and the nongaussian small scale modes add a lot of power
to the power spectrum.}
\includegraphics{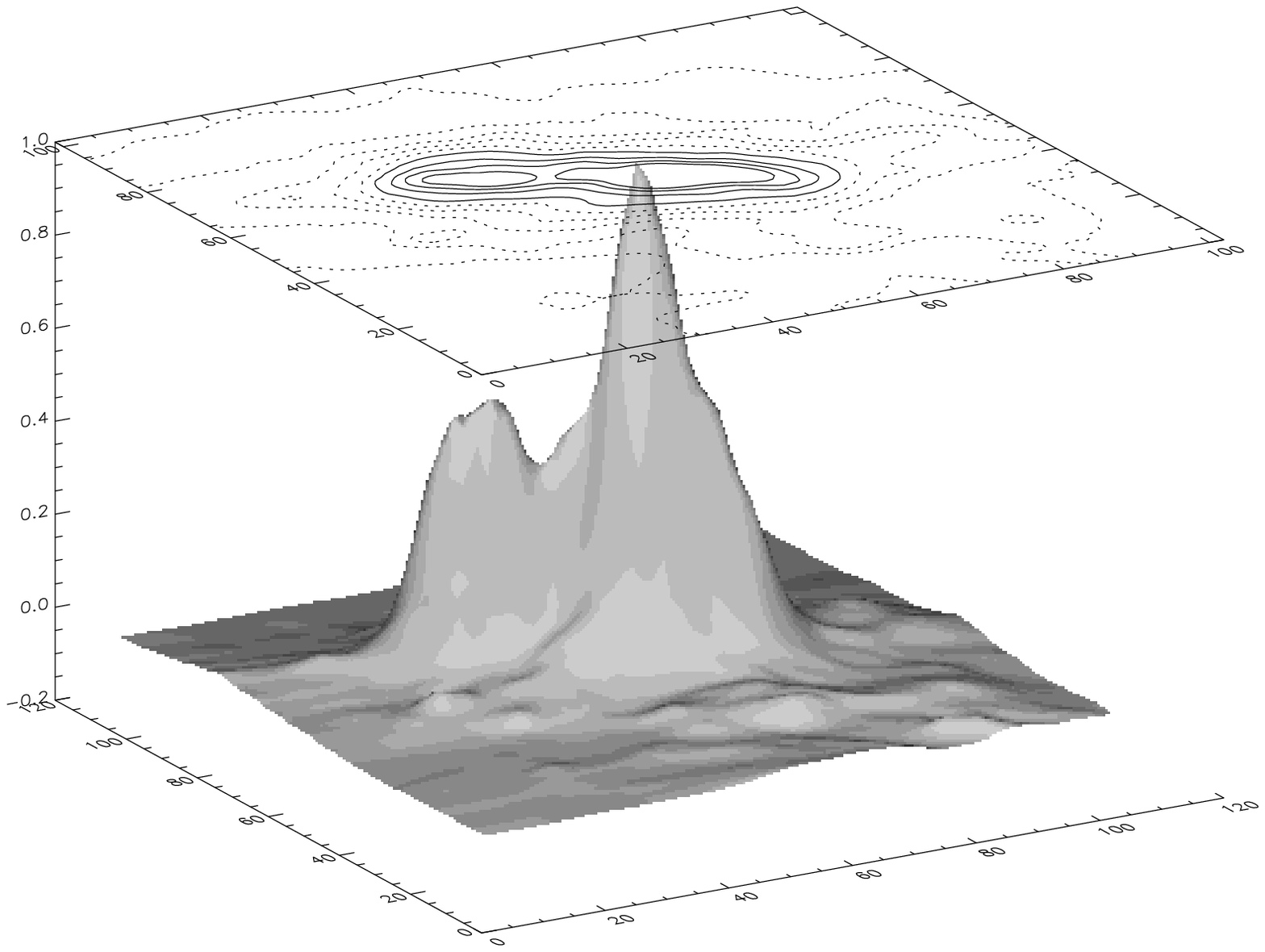}
\includegraphics{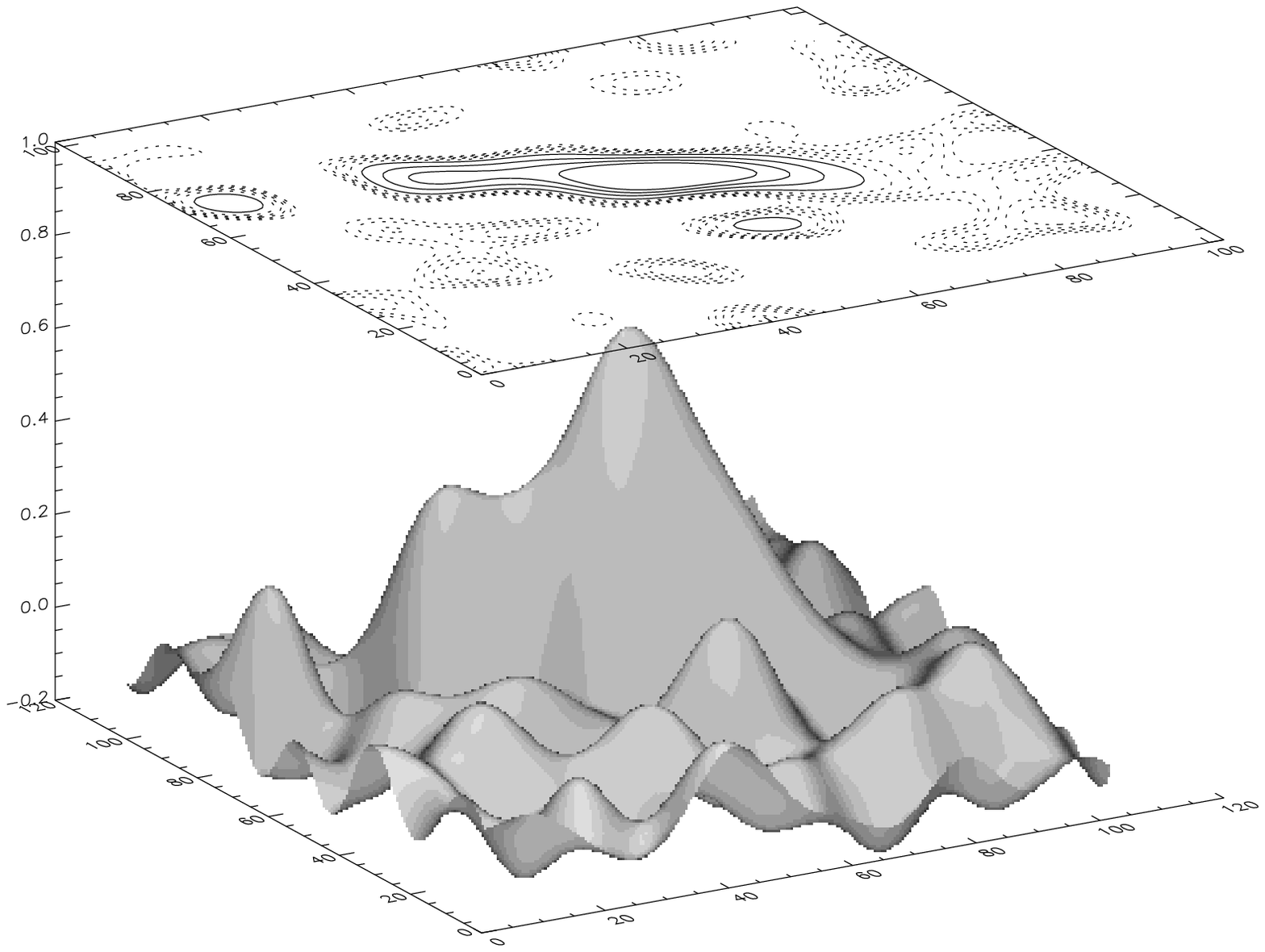}
\label{fig3_2}
\end{figure}

WF reconstruction is 
also not unique and changes with the size of the observed area. 
If we reduce the size 
then the cluster becomes more important and the contribution from
nongaussian modes to the power spectrum can dominate
over the noise to smaller scales than in previous example, 
resulting in keeping more small scale modes in  
the reconstruction. An example of this is shown in 
figure \ref{fig3_2}, where we increased the physical size of the cluster
by a factor of two, keeping the other parameters unchanged.
This enhanced the signal power spectrum and added 
the small scale structure to the reconstruction. Most of this structure
outside the cluster 
is noise, but now this noise does not dominate the power spectrum and 
the minimum variance estimator is allowed to keep this structure and
reconstruct the true structure in the center as well. 
It is clear from this discussion that WF is no longer optimal for 
every application and can only be treated as one example 
in a wider class of linear filters. 
For a conservative reconstruction of the whole measured area WF will in
general outperform other methods. 
Other low-pass filters may however 
be more appropriate for more specific applications. 
We may conclude that
using the measured power spectrum from the data in WF is
a useful starting procedure, which should be supplemented by
other linear filters by varying the prior power spectrum. 
Nonlinear methods,
such as maximum entropy method, may provide an even 
better reconstruction of strongly nongaussian and point-like structures. 

\subsection{Nonlinear effects}
So far we ignored the term $1-\kappa$ in relating observed ellipticity
to underlying shear field by assuming that 
$\kappa$ is small. When the clusters are nearly critical, 
such as in the core of the
cluster we used above, the corrections become important and have to
be included. 
Because the final result of the reconstruction 
is convergence $\kappa$ one can include the correction using
an iterative scheme: one first reconstructs $\kappa$ using equation 
(\ref{wf}) ignoring the 
correction and in the next step one uses the value of reconstructed 
$\kappa$ to obtain the components of the shear $\gamma_i$
from the observables $e_i$
\begin{equation}
\gamma_i(\bi{\theta})=e_i(\bi{\theta})/[1-\hat{\kappa}(\bi{\theta})].
\label{ge}
\end{equation}
These corrections tend to suppress the reconstructed 
convergence, because the shear is reduced for the same measured ellipticity. 
A similar iterative scheme has been proposed by \cite{ss2}.
Assuming that $\hat{\kappa}(\bi{\theta})$ is not close to unity then the error 
on the shear will continue to be dominated by the error on ellipticity
$e_i$ and one can repeat the analysis as before with new values of 
the shear. Otherwise one can improve this by adding additional
term to the error matrix 
that corresponds to the error contribution from the reconstructed
convergence, which is given by (equation \ref{var})
\begin{equation} \langle
[\hat{\kappa}(\bi{\theta})-\kappa(\bi{\theta})]
[\hat{\kappa}(\bi{\theta})-\kappa(\bi{\theta})]^{\dag}\rangle
=\bi{R}_{\kappa}\bi{D}^{-1}\bi{R}_{\kappa} 
\end{equation}
and properly add the two error contributions. If the cluster contains
a critical line where $|\det \Phi|=0$ 
then inside outer line (but outside
inner critical line) one should replace $e_i$ with $e_i/e^2$ 
(\cite{Kaiser95}) and one way to handle this case is 
to identify the critical lines at the previous iteration,
replace $e_i$ with $e_i/e^2$ where necessary and repeat the iteration until 
the convergence. We implicitly assumed that the 
zero point of convergence is known, which is something that cannot
be reconstructed from the ellipticity measurements alone.
When reconstructing LSS this will not be a major 
problem, since the mean convergence fluctuates by a few percent at 
most (the power from the modes larger than the size of the field
is small). One can thus impose the 
mean convergence to vanish across the survey by
setting the unreconstructed 
$k=0$ mode to zero. For cluster reconstruction it is
more appropriate to set the zero point so that 
the convergence at the outskirts of the cluster is zero and this is 
somewhat more problematic when the cluster extends to the edge of the 
field. For the example we used in this paper this 
is not so much of a problem, as one can separate the cluster from the 
surrounding area where the mean convergence should be small. One
can in fact completely avoid this problem by using a local relation 
between derivatives of shear and convergence (\cite{Kaiser95,ss}), 
although it does not appear straight-forward
to implement this in the methods developed here.

\subsection{Redshift distribution of the sources}
In section 3 we discussed how to handle the redshift distribution of the
sources for the power spectrum estimation. One approach was to divide
the galaxies into bins localized in distance and perform the analysis 
on each of these. The same approach should also 
be used for LSS reconstruction, because the deflections come from a 
broad region along the line of sight and can only be determined in a 
statistical sense. 
In the case of cluster reconstruction
the situation differs from LSS because the deflector is in a single 
lens plane and the distance to it is assumed
to be known. Instead of using convergence $\kappa$,
which depends on the lens and source positions, 
it is better to model directly the surface
density $\Sigma$, which is a 
physical quantity independent of the redshift distribution. It is
related to the convergence via the relation
\begin{equation}
\kappa={\Sigma \over \Sigma_{{\rm crit}}},\,\,\, \Sigma^{-1}_{{\rm crit}}(\chi_s,\chi_l)=
4\pi G a(\chi_l){r(\chi_l)r(\chi_s-\chi_l) \over r(\chi_s)},
\end{equation}
where $\chi_s$ and $\chi_l$ are the radial comoving distances to
the source and lens, respectively.
When the sources are all at the same distance one can work with $\kappa$ and
then obtain $\Sigma$ using the above relation. When the sources are not
at the same distance then $\kappa$ changes with $\chi_s$ for the same
surface density $\Sigma$. 
The observable $e_i$ can be expressed directly in terms of $\Sigma$ in 
analogy to equations (\ref{e}), (\ref{ge}) as
\begin{equation}
e_i(1-\Sigma/\Sigma_{{\rm crit}})\Sigma_{{\rm crit}}=\bi{R} \tilde{\bi{\Sigma}} +\bi{n}.
\label{ecrit}
\end{equation}
Because the galaxy distance is not known $\Sigma_{{\rm crit}}$ is
a stochastic variable and can be treated as the estimate plus additional 
noise term.
Assuming each galaxy has a known probability distribution of
distances $W_i(\chi)$ (in the absence of any distance information this
is just the overall galaxy distribution), we can define the
mean critical density 
\begin{equation}
\bar{\Sigma}^{-1}_{i,{\rm crit}}=\int d\chi_s W_i(\chi_s) \Sigma^{-1}_{{\rm crit}}(\chi_s,
\chi_l)
\end{equation}
and the dispersion around it
\begin{equation}
\sigma^2_i(\Sigma^{-1}_{{\rm crit}})=\int d\chi_s W_i(\chi_s) 
(\Sigma^{-1}_{{\rm crit}}(\chi_s, \chi_l)-\bar{\Sigma}^{-1}_{{\rm crit}})^2.
\end{equation}
For each galaxy we use the mean critical density $\bar{\Sigma}_{i,{\rm crit}}$
in equation (\ref{ecrit}) as the estimate and the
dispersion $\sigma^2_i(\Sigma_{{\rm crit}})$ as an additional 
source of error in the observable 
$e_i(1-\Sigma/\Sigma_{{\rm crit}})\Sigma_{\rm crit}$. 
This additional noise term
will be negligible for clusters at low $z$ (where $\Sigma_{{\rm crit}}$
is rather insensitive to the redshift distribution of sources). 
Even for high $z$ clusters this error is
typically subdominant compared to the error on the ellipticity if the
convergence is small. The only
case when this error becomes important is for  
clusters that are close to critical, where 
the uncertainty on $\Sigma_{{\rm crit}}$ in the term
$1-\Sigma/\Sigma_{{\rm crit}}$ of equation (\ref{ecrit}) is amplified and can 
dominate over the ellipticity error, in which case we can add this error
to the overall error budget of that galaxy. An alternative treatment 
of this problem has been discussed by \cite{ss3} for the case when 
the overall probability distribution of galaxy redshifts is given.
The nice feature of our approach
is that the mean distance and its 
error can be attached to each galaxy individually, 
which can take advantage of the multicolor 
photometric information. By assigning to each galaxy its
photometric redshift and the corresponding error estimate we can compute
for each galaxy separately its critical density and the dispersion 
around it. Another example when this is important is when the measurement
errors increase for fainter and presumably more distant sources.
In this case the more nearby galaxies
carry more weight than expected from their number density and one 
can correct for this by assigning to each galaxy a mean distance 
according to its observed flux.

\section{Discussion}
In this paper we address the question of optimal power spectrum 
estimation and projected mass density reconstruction from weak
lensing data. We show that for gaussian random processes both have
a well defined answer, which are in fact related and can be obtained
from the same calculation. For the power spectrum the final 
answer is given by the unbiased estimator and its covariance matrix 
(which also acts as a window function). The estimator is 
equivalent to the maximum likelihood method, but is significantly 
faster to compute. The covariance matrix of estimators
includes the contributions from noise, sampling variance and aliasing.
It is possible to invert from the 2-d power spectrum or to compute
directly from the data the 3-d power
spectrum, which will also be unbiased and minimum variance. Finally,
if we have distance information for the galaxies based on their 
photometric properties we can solve for both the power spectrum
and its time dependence. The method can be used in the nonlinear
regime, where it remains unbiased, but one has to compute the covariance matrix of 
the estimators, which now depend also on the reduced four point correlator.
This can be either estimated from the data themselves, obtained from
N-body simulations or computed using perturbation theory or 
hierarchical scaling relations. 
It is important to realize that even though the estimators were derived
from the maximum likelihood method, the final result is always a
weighted quadratic 
average of the data, where the weighting is the inverse of
covariance matrix. This is the most expensive part of the calculation and 
can be replaced by some simpler weighting if necessary, such as uniform 
or inverse noise weighting. Even in this case 
the method will remain unbiased and will in fact remain close to optimal 
on small scales. Similarly, if one wants to parametrize the power spectrum 
with some other parameters (such as its amplitude and slope) one can 
derive unbiased 
quadratic estimators specifically tailored for those parameters
instead of going through the power spectrum. This is however not 
necessary, since in 
the process of reducing the data to the power spectrum 
no information has been lost if minimum variance estimator has been used.

Another topic of interest that
we did not discuss in detail here is the question of optimal 
measurement of higher moments. 
One attempt how to measure these has been presented by
Schneider et al. (1997) using mass aperture statistic. As mentioned in the 
introduction this statistic 
uses the tangential component of the shear to
obtain a quantity that is a scalar (as opposed
to shear which is a tensor), which is 
needed to define a nonvanishing third 
moment in real space. Alternative way to obtain a scalar quantity 
is to reconstruct the convergence
in Fourier space as done in this paper. 
Within the context of the methods presented here 
the general strategy for determining the higher moments is in fact 
similar to the one leading to the minimum variance power spectrum methods,
at least in the limit where the higher moments are small. 
One first multiplies the measurements
with the inverse of correlation matrix.
This is just a consequence of
inverse variance weighting: if a given measurement has a large 
measurement error or it is strongly correlated with neighbouring
points then it has to be downweighted, because it does not add
new information to the data. 
On small scales this weighting reduces to the
simple uniform weighting as discussed in section 2. 
One then projects the data to Fourier space to make a scalar quantity.
Because we reconstruct 
the Fourier modes first one can attempt to measure higher order moments
directly
in Fourier space (e.g. bispectrum for the case of three-point statistic) 
and this has some nice advantages
(e.g. \cite{scoc}). One can for example average over all triple products
$\hat{\tilde{\kappa}}(\bi{l}_1)\hat{\tilde{\kappa}}(\bi{l}_2)
\hat{\tilde{\kappa}}(\bi{l}_3)$ with $\bi{l}_1+\bi{l}_2+\bi{l}_3=0$, 
keeping  the shape of the triangle the same and then vary this as 
a function of scale and angle between the wavevectors. As for the 
case of the power spectrum one has to compute the 
window for this statistic to make
the estimator unbiased and to place errors on the estimates. Alternatively, 
one can transform $\tilde{\bi{\kappa}}$ back to real space
by applying $\bi{R}_{\kappa}$ operator to obtain a real space quantity
that is a scalar, which can then be used to measure its skewness as a function of smoothing 
radius (filtering scale). Because all the operations
are linear on the data one can obtain the covariance matrix of 
the estimates and compute the window function needed to make the 
estimator unbiased and provide error estimate. 

For projected mass density reconstruction the WF method investigated
here works best on large scales, where the distribution is close to
gaussian. In that case WF preferentially keeps the structure that 
is above the noise and filters out scales that are noise dominated.
To apply WF one
has to compute the power spectrum on the data first using the 
methods discussed above. When the structures become nonlinear 
WF remains a useful reconstruction technique and still minimizes 
the variance in the class of linear estimators, but tends to oversmooth 
the data in very dense regions such as clusters. 
In that case it should be supplemented with
linear filtering with a shorter smoothing scale or with one of 
the nonlinear methods, depending on the particular application one 
has in mind. This will be particularly important if one is searching
for clusters in a random survey, where their positions are not 
given in advance and one wants to maximize the signal to noise 
for detecting such structures. 

\acknowledgements
I thank B. Jain and U. Pen for useful discussions, P. Schneider for 
helpful comments on the manuscript and U. Pen
for providing the cluster simulation data.

\end{document}